\documentclass[%
 reprint,
 nofootinbib,
 aps,
 prb,
 citeautoscript,
 amsmath,amssymb,
]{revtex4-1}

\usepackage{amsmath}
\usepackage{graphicx}
\usepackage{dcolumn}
\usepackage{bm}
\usepackage{epsfig,color,xspace,multirow,xr,bbold}
\usepackage[all]{xy}
\usepackage{setspace}
\usepackage{url}
\usepackage{verbatim}
\usepackage{fancyvrb}

\usepackage[labeled,resetlabels]{multibib}
\newcites{SI}{SI References}

\newcommand{\dGwol}{\Delta G_{\textup{W}\rightarrow\textup{Ol}}}
\newcommand{\dGwm}{\Delta G_{\textup{W}\rightarrow\textup{M}}}
\newcommand{\pka}{p$K_{\rm a}$\xspace}
\newcommand{\apka}{ap$K_{\rm a}$\xspace}
\newcommand{\bpka}{bp$K_{\rm a}$\xspace}

\begin{document}

\author{Roberto Menichetti}
\author{Kiran H.~Kanekal}
\author{Tristan Bereau}
\email{bereau@mpip-mainz.mpg.de}
\affiliation{Max Planck Institute for Polymer Research, 55128 Mainz,
  Germany}

\title{Drug-membrane permeability across chemical space}

\begin{abstract}
 Unraveling the relation between the chemical structure of small
 drug-like compounds and their rate of passive permeation across lipid
 membranes is of fundamental importance for pharmaceutical
 applications.  The elucidation of a comprehensive
 structure-permeability relationship expressed in terms of a few
 molecular descriptors is unfortunately hampered by the overwhelming
 number of possible compounds.  In this work, we reduce a priori the
 size and diversity of chemical space to solve an analogous---but
 smoothed out---structure-property relationship problem.  This is
 achieved by relying on a physics-based coarse-grained model that
 reduces the size of chemical space, enabling a comprehensive
 exploration of this space with greatly reduced computational cost.
 We perform high-throughput coarse-grained (HTCG) simulations to
 derive a permeability surface in terms of two simple molecular
 descriptors---bulk partitioning free energy and \pka.  The surface is
 constructed by exhaustively simulating all coarse-grained compounds
 that are representative of small organic molecules (ranging from 30
 to 160 Da) in a high-throughput scheme.  We provide results for
 acidic, basic and zwitterionic compounds.  Connecting back to the
 atomic resolution, the HTCG predictions for more than 500,000
 compounds allow us to establish a clear connection between specific
 chemical groups and the resulting permeability coefficient, enabling
 for the first time an inverse design procedure.  Our results have
 profound implications for drug synthesis: the predominance of
 commonly-employed chemical moieties narrows down the range of
 permeabilities.
\end{abstract} 

\maketitle

\section{Introduction}

\begin{figure*}[htbp]
  \begin{center}
    \includegraphics[width=0.95\linewidth]{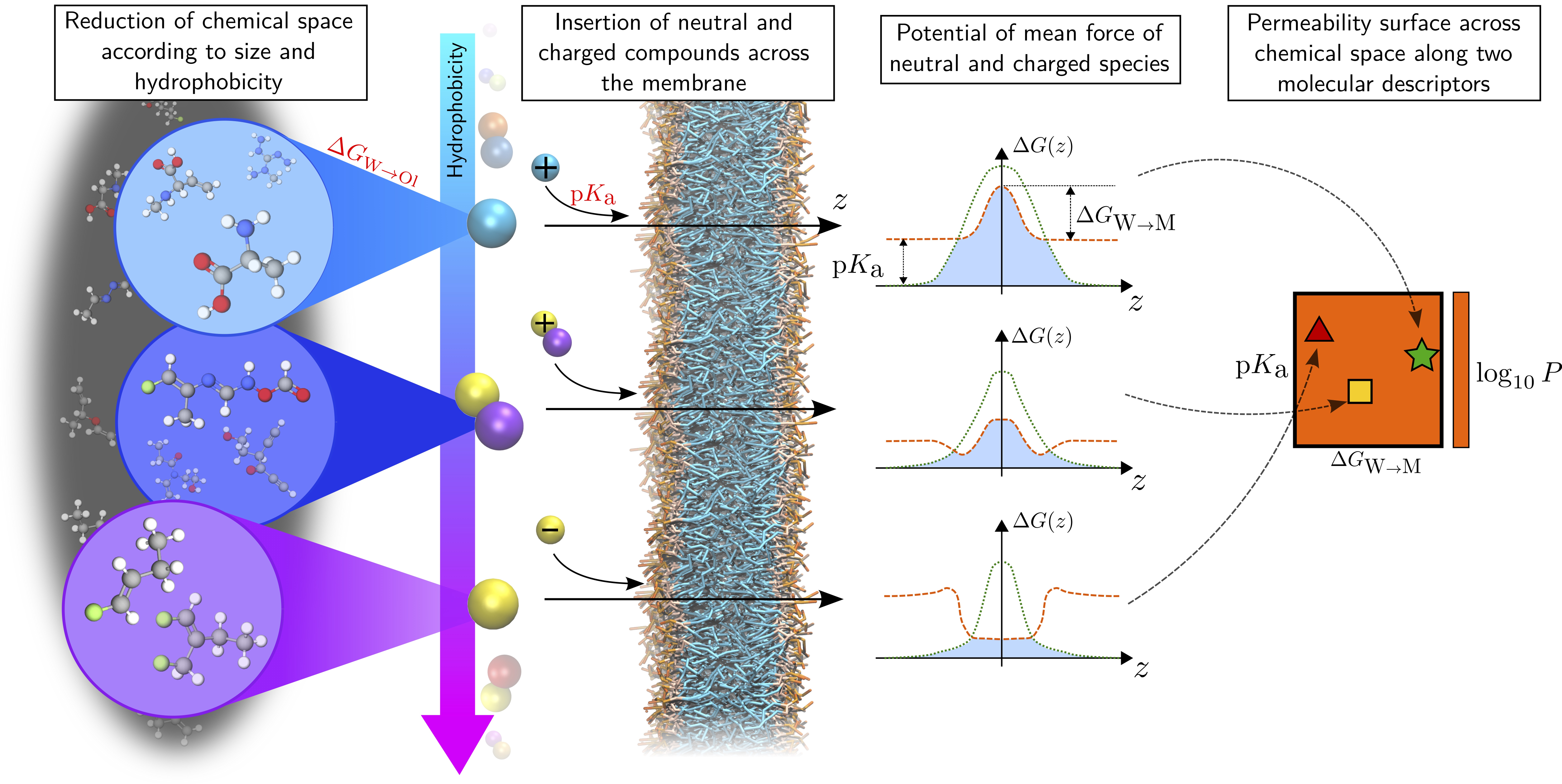}
    \caption{From left to right: Coarse-graining reduces the size of
      chemical space, such that many small molecules of similar size
      and hydrophobicity get mapped to the same
      representation.\cite{menichetti2017silico} For each molecule, we
      model its passive translocation across a lipid bilayer (water
      not shown for clarity).  The thermodynamics of the system is
      characterized by the potential of mean force (PMF), evaluating
      both the neutral and charged species, shifted according to the
      compound's \pka.  The major dependence of the PMF on the
      water/octanol partitioning and the \pka motivates these as
      molecular descriptors to construct a permeability surface
      (Eq.~\ref{eq:perm}).  These two molecular descriptors, also
      highlighted in red, are experimental quantities directly fed
      into the physics-based simulations to yield a parameter-free
      estimation of the permeability coefficient.  }
    \label{fig:intro}
  \end{center}
\end{figure*}

The passive permeation of small molecules across lipid membranes
offers not only physico-chemical insight but also crucial
pharmaceutical information about drug-membrane
thermodynamics.\cite{lipinski2001experimental} It probes the timescale
of translocation due to a concentration gradient of the drug, without
active cellular mechanisms (Fig.~\ref{fig:intro}).  A detailed
understanding of the underlying structure-property relationships
between drug chemistry and passive-permeation thermodynamics, though
of great interest for drug development, is still lacking.

Structure-property relationships are often tackled by means of
high-throughput screening experiments: ($i$) a large number of
compounds are probed with respect to the property of interest by
individual measurements or calculations; ($ii$) the relationship
between structure and property is empirically learned by means of
statistical algorithms.\cite{colon2014high, di2015drug, Bereau2016}
While structure-property relationships thus formally rely on both the
breadth and quality of the data, as well as the accuracy of the
statistical model, the common bottleneck in the pharmaceutical
sciences often arises from the former.

Even though in vivo techniques probe drug-membrane interactions in all
the intricacies of the cellular environment, the experimental cost and
complexity make them poorly suited for high-throughput
screening.\cite{pardridge1995transport} It is instead the development
of in vitro techniques that have helped in expanding
passive-permeation databases.\cite{pidgeon1995iam,
  yazdanian1998correlating} Unfortunately, limited aggregate data has
been made publicly available thus far.  The resulting statistical
models---such as quantitative structure-property relationships (QSPR)
or machine learning---typically rely on $10^2$ to $10^4$ datapoints
only.\cite{stouch2003silico, swift2013back, zhang2017machine} The
question follows: how representative can these samples be, when the
size of drug chemical space is estimated at
$10^{60}$?\cite{Dobson2004} The tendency of these statistical models
to depend significantly on individual outliers strongly suggest
overfitting---these models lack transferability across chemical space.
\cite{swift2013back} The small-dataset problem is typically aggravated
by the compounds' poor diversity.\cite{spring2003diversity}

As a complementary approach to experimental measurements,
physics-based modeling provides a robust strategy to predict passive
permeation in silico.\cite{orsi2010passive, swift2013back} The
inhomogeneous solubility-diffusion
model\cite{diamond1974interpretation,marrink1994simulation} considers
the concentration gradient of a solute molecule across an interface to
yield a permeability coefficient, $P$.\cite{votapka2016two} This
results in a spatial integral normal to the interface, $z$, of the
potential of mean force (PMF), $G(z)$, and local diffusivity, $D(z)$,
\begin{equation}
  \label{eq:perm}
  P^{-1} = \int {\rm d}z \frac{\exp[\beta G(z)]}{D(z)},
\end{equation}
with $\beta=1/k_{\text{B}}T$. Eq.~\ref{eq:perm} highlights the two key
parameters that contribute to the rate of passive permeation of a
compound: its hydrophobicity, quantified by the PMF, together with the
local diffusivity.  Practically, $G(z)$ and $D(z)$ are commonly
extracted from enhanced-sampling classical molecular dynamics
simulations.  Grounding the problem within the statistical mechanics
of a concentration flux diffusing through an interface combined with
conformational sampling from physically-motivated force fields can
offer unprecedented insight.  We stress that current experimental
techniques have yet to resolve $G(z)$---computer simulations thus
remain the gold standard to estimate Eq.~\ref{eq:perm}.
Unfortunately, adequate conformational sampling remains
computationally daunting at the atomistic level, even for a small
rigid molecule crossing a single-component lipid membrane: roughly
$10^5$~CPU-hours per compound limit this strategy to up to $\sim10$
different molecules per study.\cite{carpenter2014method,
  lee2016simulation, bennion2017predicting,tse2018link} When combined
with the overwhelming size of chemical space, these figures hinder
short-term prospects of running atomistic simulations at high
throughput, thereby hampering the elucidation of the underlying
structure-property relationships.

Structure-property relationships effectively project down chemical
complexity on a few molecular descriptors that map to the property of
interest.\cite{ghiringhelli2015big, isayev2017universal} Inferring
these maps typically relies on a statistical analysis over many
measurements, identifying a \emph{smooth} (i.e., low-dimensional)
connection between structure and property.  In this work we propose an
alternative strategy: rather than smoothing this connection a
posteriori, we enforce it a priori.  We still rely on physics-based
models but reduce their resolution to efficiently interpolate across
chemistry, while ensuring accurate thermodynamics by construction.
This enables a high-throughput approach for two reasons:
($i$) the reduced representation significantly speeds up every
simulation, and ($ii$) the interpolation across chemistry effectively
reduces the size of chemical space.  Solving the structure-property
relationship problem for the reduced model proves significantly more
tractable, and allows the identification of a permeability surface as
a function of simple molecular descriptors. We further connect back to
the original problem by means of a large-scale analysis of our
predictions.

The abovementioned reduced models, better known as coarse-grained (CG)
models, lump together several atoms into a bead.\cite{Noid2013,
  voth2008coarse} While defined in terms of fewer degrees of freedom,
coarse-graining remains physics-based and can be combined with
rigorous free-energy calculations.  Here, we rely on the CG Martini
model, which has shown useful to simulating a wide variety of
biomolecular systems.\cite{M-2004,M-2007,marrink2013perspective} In
Martini, a small set of bead types encodes how small organic fragments
partition between solvents of different polarity, thus ensuring robust
thermodynamics at complex interfaces,\cite{marrink2013perspective}
while cutting down the computational costs by three orders of
magnitude.\cite{menichetti2017efficient}

The parametrization of a molecule at the CG level thus consists of a
collection of Martini beads, each representing a specific chemical
group.  Constructing a CG molecule can be streamlined into a
systematic procedure,\cite{Bereau2015} so as to emulate the molecule's
overall shape and hydrophobicity.  The small set of bead types leads
to a degeneracy in the representation: many molecules of similar
shapes and hydrophobicity map to the same CG parametrization
(Fig.~\ref{fig:intro}).  Such a many-to-one mapping generates a
significant reduction in the size of chemical space---further lowering
the computational investment by an additional $10^3-10^4$.  The few
bead types involved leads to a dramatic reduction in the combinatorial
explosion of chemistry, easing the construction of \emph{all} CG small
molecules up to a certain size
(Fig.~\ref{fig:intro}).\cite{menichetti2017silico} This addresses the
poor-diversity issues that synthetic databases typically face,
facilitating a representative coverage of subsets of chemical space
projected primarily along size and hydrophobicity.  Recently, these
properties allowed us to predict the PMF of drug-membrane partitioning
for an unprecedented $511,427$ small molecules---several orders
of magnitude beyond what was previously
available.\cite{menichetti2017silico} In terms of accuracy we showed
that the CG model achieves a mean-absolute error of 0.8~kcal/mol to
predict bulk water/octanol partitioning free energies,\cite{Bereau2015} 
translating to a 1.4~kcal/mol error along a PMF.\cite{menichetti2017silico} 
We further stress that these errors are
evaluated across a significant subset of the chemistry of small
organic molecules, while atomistic results are too scarce to make such
estimates.  For the present work, our error estimates roughly
translate to an accuracy of $1 \log_{10}$ unit in the permeability
coefficient, validated across an extensive set of
structurally-distinct compounds against both atomistic simulations and
experimental measurements (see Supporting Information (SI)).

In this work, we use high-throughput coarse-grained (HTCG) simulations
to cover a subset of chemical space both efficiently and broadly.
Unlike conventional high-throughput screening protocols that require
an arbitrary selection of compounds, we consider \emph{all}
coarse-grained representations up to a threshold size, mapping to most
small organic molecules ranging from 30 to 160 Da.  This comprehensive
exploration allows us to systematically investigate the effect of
hydrophobicity and \pka on the permeation rate (Fig.~\ref{fig:intro}),
unlike previous studies limited to a handful of
compounds.\cite{carpenter2014method, lee2016simulation,
  bennion2017predicting,tse2018link} Our methodology offers a unique
approach to construct a two-dimensional surface describing the
permeability of a small molecule across a lipid membrane
(Fig.~\ref{fig:intro}). The molecular descriptors are here motivated
by the physics of the permeation process, i.e., the interplay between
diffusivity and solubility (Eq.~\ref{eq:perm}).  Because the
diffusivity was shown to be rather insensitive to chemical
detail,\cite{carpenter2014method} we focus on the potential of mean
force, $G(z)$.  We have recently shown that the key features of $G(z)$
can be reconstructed simply from the bulk-partitioning free
energy.\cite{menichetti2017silico} By further accounting for the
contribution of different protonation states, we also express the
permeability surface in terms of its acid dissociation constant in
water, \pka.  In the following we focus on acidic and basic compounds,
while the SI further discusses zwitterions.  These surfaces allow for
a rapid, \emph{simulation-free} prediction of drug permeability
starting from key molecular properties.  The accuracy is roughly on
par with explicit CG simulations due to compensating errors between
the two methods.

Extracting permeability surfaces from the CG simulations allows us to
connect back to the original structure-property relationship problem.
Our analysis of over 500,000 small molecules mapping to the
investigated CG representations unveils the role played by
representative functional groups in the permeability coefficient,
enabling inverse molecular design.  The link drawn here has profound
implications for drug synthesis: favoring the incorporation of certain
chemical groups (e.g., carboxylic groups) will reduce the range of
accessible permeabilities of the final compound.

\section{Results and discussion}

While drug permeation is known to depend on lipid
composition,\cite{tse2018link} in this work we only consider a
single-component bilayer made of
1,2-dioleoyl-\emph{sn}-glycero-3-phosphocholine (DOPC).  The
permeability coefficient, $P$, is readily estimated from the PMF and
diffusivity profile (Eq.~\ref{eq:perm}).  The PMFs are extracted from
HTCG simulations of \emph{all} CG representations made of one and two
beads, mapping to a representative subset of small organic molecules
in the range $30-160$~Da.\cite{menichetti2017silico} For compounds
capable of (de)protonating, we also model the corresponding charged
species.  For convenience, we distinguish the \pka of a chemical group
as being either acidic (\apka) or basic (\bpka), which quantifies the
propensity of a \emph{neutral} compound to deprotonate or protonate,
respectively.  The effective permeability coefficient is constructed
by a combination of the two PMFs (Fig.~\ref{fig:intro}), shifted
according to the compound's \pka in water, see
Methods.\cite{maccallum2008distribution, marvin} The diffusivity
profile is estimated from reference atomistic
simulations.\cite{carpenter2014method}

\subsection{Permeability surfaces}

\begin{figure}[!htp]
  \begin{center}
    \includegraphics[width=0.85\linewidth]{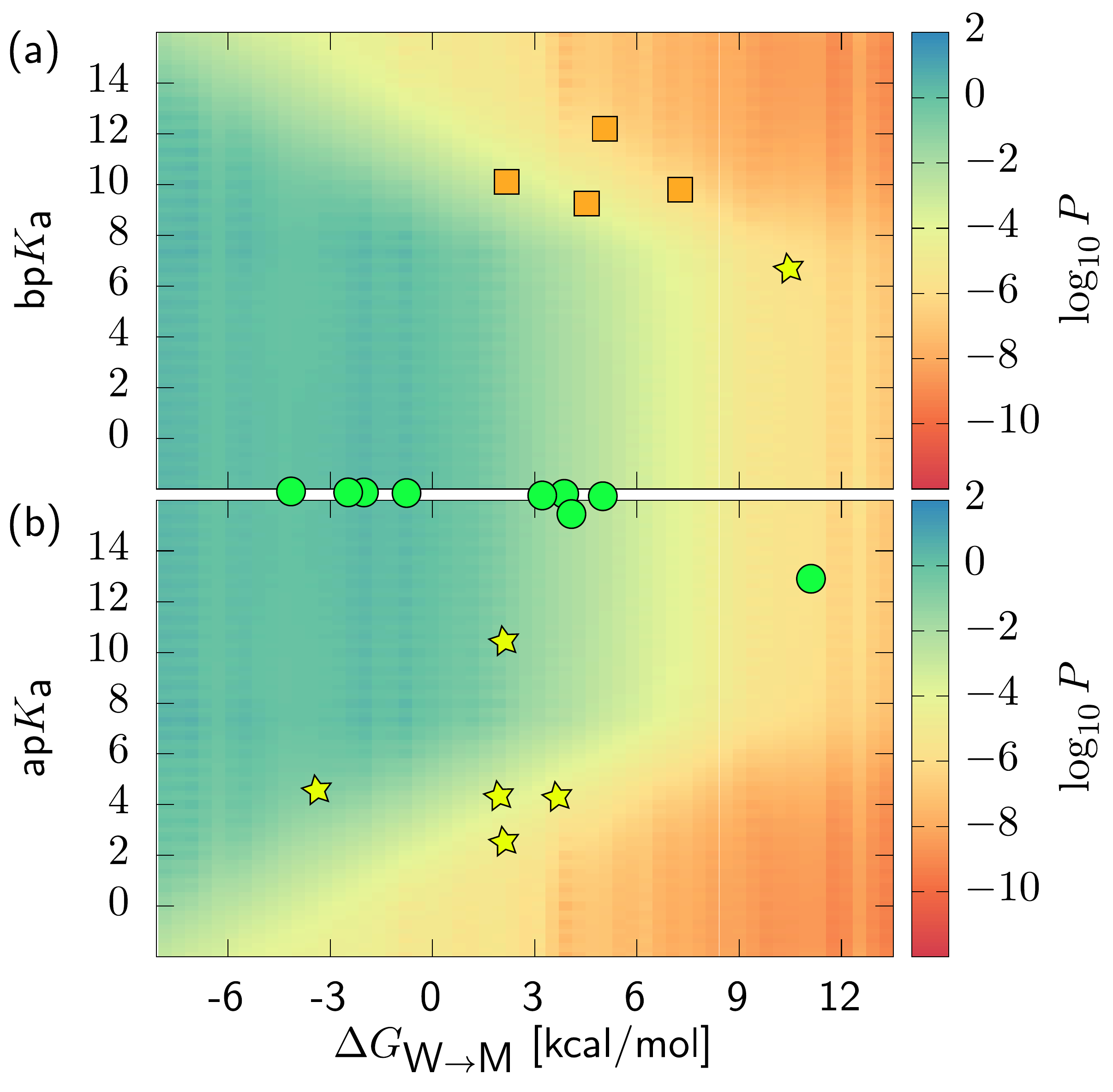}
    \caption{Permeability surfaces ($\log_{10}$ scale) calculated from
      HTCG simulations as a function of two small-molecule
      descriptors: the (a) basic or (b) acidic \pka in water and the
      water/membrane partitioning free energy, $\dGwm$.  Cooler
      (warmer) colors correspond to faster (slower) permeating
      molecules.  The intersection between the two surfaces
      corresponds to compounds that effectively always remain neutral.
      Green circles, yellow stars, and orange squares correspond to
      deviations from atomistic simulations within 0.5, 1.3, and 2.2
      log units, respectively (SI). }
    \label{fig:permsurf}
  \end{center}
\end{figure}

Fig.~\ref{fig:permsurf} displays the computed drug-membrane
permeability as a function of two drug parameters: its \pka in water
and water/membrane partitioning free energy, $\dGwm$.  The latter
corresponds to the free energy difference between insertion in bulk
water and the membrane-bilayer midplane.  Though we have shown this
quantity to correlate extremely well with the
experimentally-accessible water/octanol partitioning free energy,
$\dGwol$, $\dGwm$ displays enhanced transferability across CG
molecular sizes.\cite{menichetti2017silico} Indeed, HTCG simulations
of single-bead or two-bead CG compounds lead to identical permeability
surfaces, except for the range of $\dGwm$ covered (compare Fig.~S4
with Fig.~\ref{fig:permsurf}).

Fig.~\ref{fig:permsurf} displays smooth permeability surfaces as a
function of the drug's acidic and basic \pka value in water.  The
$\log_{10}$ scale of the permeability surfaces indicates the wide
timescale variations these molecular parameters exert on the
thermodynamic process.  For both panels, the horizontal behavior
indicates that larger permeabilities are obtained toward the
left---more hydrophobic compounds---while polar molecules experience
more difficulties crossing the lipid bilayer, leading to a drastic
reduction in $P$.  The effect is compounded by (de)protonation: panel
(a) across the vertical axis describes the effect of the compound's
\apka in water onto $P$.  Extremely strongly acidic molecules (\apka
$ \lesssim 2$) effectively remain charged across the membrane
interface, leading to prohibitively large free energies along the PMF,
such that their rate of permeation is strongly suppressed.  Increasing
\apka shows a significant increase in $P$, up to \apka $\approx 7$,
beyond which $P$ plateaus.  This stabilization is due to the
competition between neutral and charged PMFs, where the charged PMF is
shifted to increasingly larger values, and therefore never contributes
significantly compared to the more attractive neutral PMF.  Of
particular interest are the strong acids ($2 \lesssim $ \apka
$ \lesssim 7$), which neutralize upon entering the membrane,
effectively enhancing the permeability coefficient as compared to a
compound that remains charged across the interface.  An approximately
symmetric behavior can be observed when switching from acidic to basic
compounds (panel (b)).  The impact of both \apka and \bpka on the
permeability coefficient becomes even more pronounced in the case of
zwitterions (Fig.~S5), where high permeation rates are only obtained
for compounds containing both weak acidic \emph{and} basic chemical
groups.

The permeability surface also displays a comparison against atomistic
simulations\cite{maccallum2008distribution, carpenter2014method,
  lee2016simulation} for several compounds (symbols in
Fig.~\ref{fig:permsurf}).  These points provide a validation of our
methodology---we report a mean absolute error of $1.0~\log_{10}$ unit
across the two molecular descriptors---with additional information
included in the SI (also against experimental data).  Most
importantly, the few datapoints highlight the extremely limited
exploration of chemical space using \emph{in silico} simulations at an
atomistic resolution.

\subsection{Functional-group localization on the permeability
  surfaces}

To better elucidate how the chemical structure impacts the
permeability coefficient, we consider a large database of small
organic molecules from combinatorial chemistry: the generated database
(GDB).\cite{Fink2005,Fink2007} It consists of a large set of stable
molecules up to 10 heavy atoms made of the chemical elements C, O, N,
and F, saturated with H.  We pointed out how transferable
coarse-grained models effectively reduce the size of chemical space by
lumping many molecules into one coarse-grained
representation.\cite{menichetti2017silico} This allows us to associate
the abovementioned one- and two-bead CG permeability results to $5
\times 10^5$ molecules.  The distinction made between compounds that
reduce to CG molecules made of a single bead (``unimers'') from those
made of two beads (``dimers'') effectively amounts to a segregation between
molecular weights.\cite{menichetti2017silico} We populate the
permeability surfaces with these compounds---projecting them onto the
two molecular descriptors: \pka and water/octanol partitioning free energy $\dGwol$.
By coarse-graining every single compound, we establish a map between
chemical structure and its CG thermodynamic property.

\begin{figure*}[htbp]
  \begin{center}
    \includegraphics[width=0.7\linewidth]{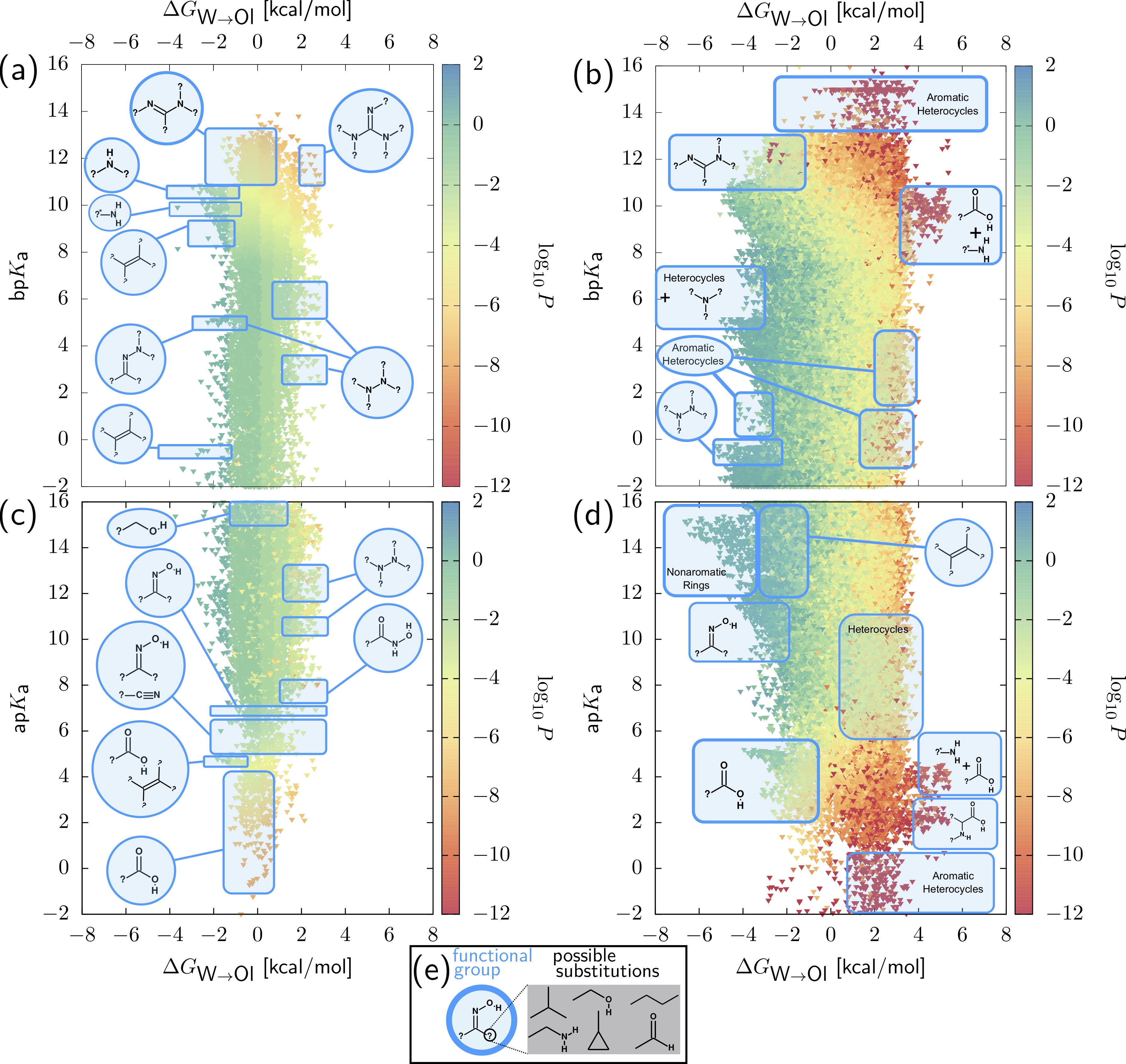}
    \caption{Chemical-space coverage of GDB projected onto \pka and
      water/octanol partitioning free energies, $\dGwol$.  Basic and
      acidic \pka are shown in panels (a,b) and (c,d), respectively.
      Panels (a,c) and (b,d) describe the coverage corresponding to
      coarse-grained unimers and dimers, respectively.  Regions
      highlighted in light blue display several representative
      chemical groups.  Substitutions denoted by ``{\bf ?}''
      correspond to either H or a substitution starting with an alkyl
      or aryl carbon, while ``{\bf ?*}'' only corresponds to
      substitutions that begin with an alkyl carbon.  (e) Our analysis
      clusters molecules containing both a predominant functional
      group (blue), but also one or several substitutions (black), of
      which only a few possibilities are shown. }
    \label{fig:fgroup}
  \end{center}
\end{figure*}

Fig.~\ref{fig:fgroup} displays the chemical-space coverage of GDB
compounds onto the molecular descriptors. For all panels, we have
colored the points in terms of the permeability calculated using HTCG
simulations.  Top and bottom panels distinguish between \bpka and
\apka, while left and right denote unimers and dimers, respectively.
We first note that the cloud of points is not uniformly distributed,
but is instead centered around zero in $\dGwol$.  An increase in the
molecular weight of the compound (left to right in
Fig.~\ref{fig:fgroup}) opens up new regions of chemical space, as we
observe a significant broadening of the distribution along the
water/octanol axis. This naturally arises due to the extensivity of
the water/octanol partitioning, the more complex combinatorics of
atoms involved, and the additional presence of five-membered rings.

Unlike bulk partitioning, the \pka of a compound is not significantly
impacted by aggregate behavior, but is instead dominated by one or a
few specific chemical groups capable of (de)protonating.  As such, we
investigated the presence of chemical groups representative of a
subset of chemical space.  The regions in blue highlight a chemical
group that is predominant, appearing in at least 50\% of the molecules
in that subset.  Detailed statistics pertaining to the frequency of
specific functional groups in each of the blue regions is provided in
the SI.  The localization of chemical groups remains largely similar
from unimers to dimers (e.g., carboxylic
group).  Our high-throughput analysis offers an intuitive
visualization of the link between chemistry and permeabilities via the
\pka.  Fig.~\ref{fig:fgroup} reflects that oxygen-containing
functional groups are generally more likely to be proton donors,
whereas nitrogen-containing functional groups can serve as either
proton donors or acceptors.\cite{bruice2016organic} At low \apka
values, we mainly see carboxylic groups transitioning to
nitrogen-containing functional groups (e.g., oxime derivatives) as we
increase the \apka.  Contrastingly, the \bpka chemical coverage
displays no predominant oxygen-containing functional groups.  Notable
exceptions are the zwitterionic amino acid-like compounds and certain
aromatic heterocyclic compounds shown
in Fig.~\ref{fig:fgroup}, which have both a low \apka and a high
\bpka.  These functional groups largely contribute to the chemical
coverage of zwitterions (Fig.~S5b).

\subsection{Linking functional groups and the permeability surface
  enables molecular design}

Fig.~\ref{fig:fgroup} enables a robust ad hoc method for both direct
and inverse molecular design.  The direct route amounts to estimating
the permeability coefficient given a chemical structure.
Fig.~\ref{fig:fgroup} simply requires an estimate for the two
molecular descriptors, \pka and $\dGwol$, either from experiments or
prediction algorithms.\cite{Tetko2002, marvin} More interestingly, our
results allow us to focus on specific regions of chemical space
compatible with a desired permeability coefficient.  We effectively
reduce the high dimensionality of chemical space by projecting down
onto our molecular descriptors and identifying key scaffolds.

Fig.~\ref{fig:fgroup} offers a simple route at an inverse design
procedure.  For example, if designing a small molecule of 3 to 5 heavy
atoms (i.e., mapping to a CG unimer) that requires a $\log_{10} P$ of
$-1.0$, Fig.~\ref{fig:fgroup}c suggests molecules containing either a
terminal hydroxyl group or an oxime group.  Indeed, small alcohols
such as propanol and butanol match this target (Fig.~S8), although we
are not aware of relevant experimental studies containing small oxime
derivatives.  Interestingly, we can also predict how small chemical
changes will affect permeability: a change that impacts hydrophobicity
(e.g., through hetereoatom substitions) will smoothly shift the
compound horizontally on the surface.  On the other hand, the
introduction of new (de)protonatable groups might lead to large jumps
on the surface, dictated by the strongest acid or base present in the
molecule.  The different behavior across the horizontal and vertical
axes is due to the extensive and intensive characters of the
descriptors, respectively.

Critically, Fig.~\ref{fig:fgroup} shows remarkable transferability
\emph{outside} the range of compounds used in the screening. For
example, while salicylate is made up of 10 heavy atoms, its aromatic
ring leads to a four-bead representation.  CG simulations using this
parametrization result in $\log_{10} P=-4.21$ (Fig.~S6 and Table S1),
deviating only one $\log_{10}$ unit from the atomistic results
(highlighted as one of the symbols in
Fig.~\ref{fig:permsurf}).\cite{carpenter2014method} Alternatively, we
can easily read off the permeability from the surface: the carboxylic
group is the main contributor for its descriptors \apka $= 2.8$ and
$\dGwol = -2.7$~kcal/mol (Fig.~\ref{fig:fgroup}).  This results in a
\emph{simulation-free} prediction for $\log_{10} P$ of $-3.72$, less
than two log units away from the atomistic results. The discrepancy
between the four-bead representation and the dimer surface we rely on
is the main source of errors: we have observed a systematic shift
between $\dGwol$ and $\dGwm$ as a function of the number of CG
beads.\cite{menichetti2017silico} An even more challenging test case
involved ibuprofen ($206$~Da, significantly outside our range of
molecular weights), for which both CG simulations and the surface
prediction yield an accuracy within $1 \log_{10}$ unit within the
atomistic results (symbol in Fig.~\ref{fig:permsurf}, Fig.~S6, and
Table~S1).

We verified this consistent accuracy between explicit CG simulations
and simulation-free surface predictions across two dozen small
molecules---both in and out of the range of molecular weights
considered (Fig.~S7 and Table~S1).  Although one would expect higher
accuracy from explicit simulations, we observe compensating errors
between the discretization of partitioning free energies and the
smoothing of the surface.  The transferability beyond the initial
molecular weight considered speaks to the robustness of our
physics-based approach.  This feature contrasts radically with
statistical methods that fit experimental data, such as QSPR: the
transferability of a QSPR model hinges upon potential biases in the
training dataset.  Given the small dataset sizes available from
experiments and the wider range of molecular weights, QSPR models tend
to be limited to chemistries very close to those used in
training.\cite{Lambrinidis2017QSPR, Wang2016QSPR} On the other hand,
the HTCG method systematically spans a wide region of chemical
compound space without resorting to parameter tuning, offering
accurate predictions even beyond the range of molecular weight
considered.

\subsection{Impact of functional-group localization on bioavailability}

\begin{figure}[htbp]
  \begin{center}
    \includegraphics[width=0.7\linewidth]{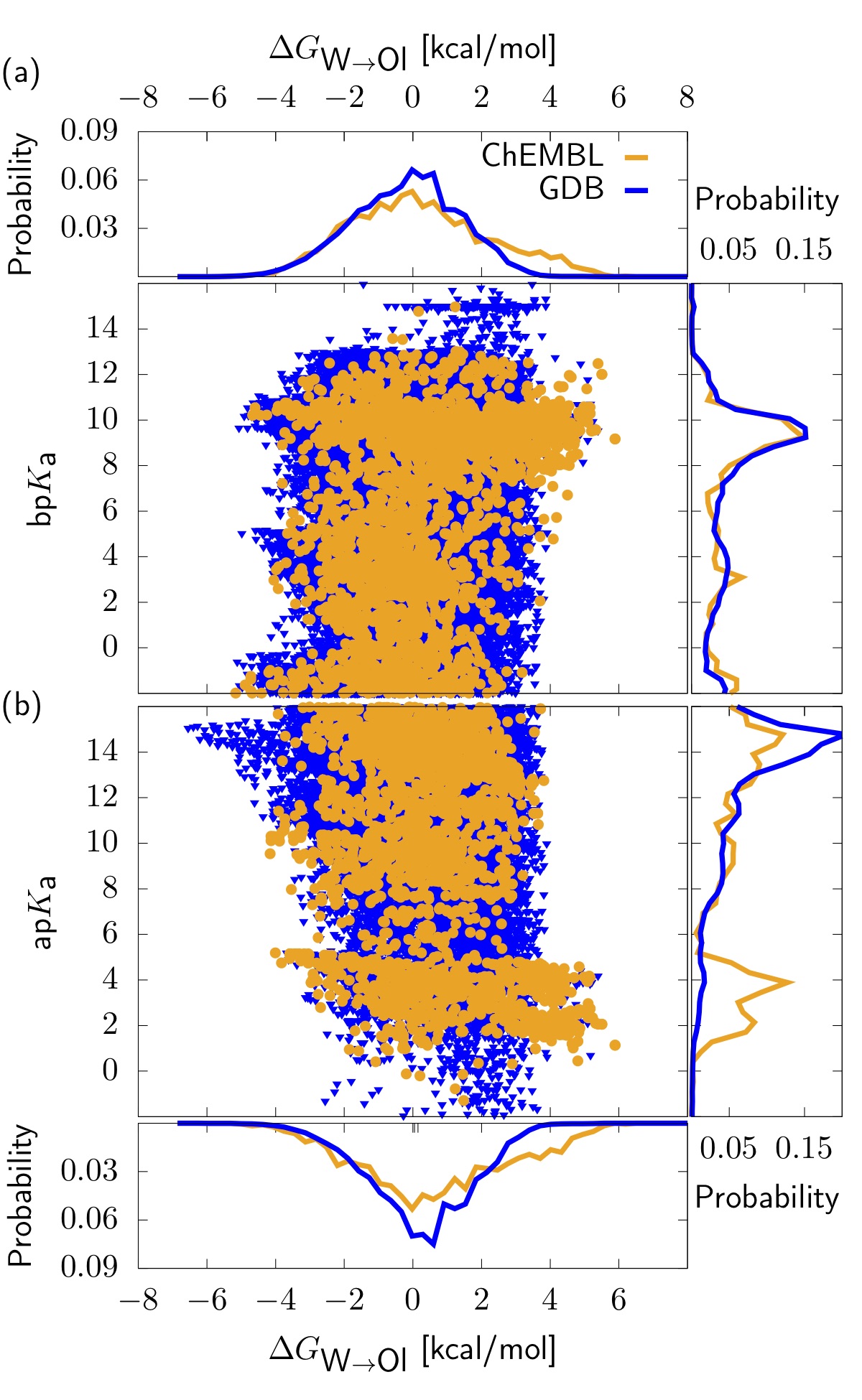}
    \caption{Comparison of the chemical-space coverage of the
      combinatorial GDB and synthetic ChEMBL databases, projected onto
      (a) basic or (b) acidic \pka and water/octanol partitioning free
      energy, $\dGwol$.  The coverages are further projected down
      along a single variable on the sides.  Note the significant
      differences between the GDB and ChEMBL distributions along the
      \apka in panel (b).}
    \label{fig:cov}
  \end{center}
\end{figure}

The projection of the GDB database onto the two molecular descriptors
provides a low-dimensional representation of chemical-space coverage.
Interestingly, this helps compare its breadth and variety with other
databases.  In particular, we focus on ChEMBL: a database of
synthesized compounds.\cite{bento2014chembl} We prune ChEMBL to only
retain compounds roughly compatible in size with the compounds in GDB
(up to 10 heavy atoms), as well as H, C, O, N, and F elements only.
Fig.~\ref{fig:cov} displays the coverage of both GDB and ChEMBL onto
the molecular descriptors.  Here again, panels (a) and (b) distinguish
acidic and basic ionizing groups.  We first note that ChEMBL displays
a much smaller number of datapoints, illustrating the minuscule ratio
of stable compounds that have been synthesized.\cite{Dobson2004}
Overall the two databases cover remarkably similar regions of this
chemical surface.  However, a projection of the distributions along
the individual axes indicates a statistically-significant difference
for \apka: synthesized compounds strikingly overrepresent compounds
with low \apka values (from 2 to 4).  We find a significant
overrepresentation of carboxylic groups in ChEMBL: 90\% of the
compounds in the range $0 < \text{\apka} < 6$ contain such a group.
This well-known bias in drug design \cite{ballatore2013carboxylic} can
readily be rationalized: Synthesizing compounds that include
carboxylic groups will offer relatively strong acidity as well as an
improved ability to hydrogen bond---a dominant interaction in most
biomolecular processes.  Our results introduce further implications:
the overrepresentation of carboxylic groups will effectively narrow
down the range of permeability coefficients. This limitation will be
further compounded by the necessity of a drug candidate to show high
aqueous solubilities, and the delicate interplay existing between
these two properties,\cite{dahan2016solubility} overall affecting the
compounds' bioavailability.

\section{Conclusions}

We present the prediction of membrane-permeability coefficients for an
unprecedented number and chemical range of small organic molecules
across a single-component DOPC lipid bilayer.  Rather than tackling
the original structure-property relationship problem head-on, we work
with physics-based reduced models that smoothly interpolate across
chemistry, thereby reducing the size of chemical space.  Critically,
we do not arbitrarily select compounds to be screened, but instead
systematically consider \emph{all} coarse-grained representations that
map to small organic molecules ranging from 30 to 160 Da.
Coarse-grained permeability predictions were extensively validated
against both atomistic simulations and experimental measurements for
structurally-diverse compounds.  The high-throughput coarse-grained
(HTCG) simulation approach used here compounds more efficient
conformational sampling and reduction in chemical space, offering an
overall speedup of $\sim 10^6$ compared to atomistic simulations.
This enables a systematic exploration of the link between chemical
structure and permeability coefficient.  To this end we construct a
smooth surface as a function of two molecular descriptors: the \pka
and water/membrane partitioning free energy, $\dGwm$.  The many orders
of magnitude covered by the surface indicate the significant impact of
the small molecule's chemistry onto the thermodynamic process.  The
surfaces illustrate how strong acids and bases limit the loss of
permeability for charged compounds. Having solved the reduced
structure-property mapping allows us to connect back to the original,
higher-dimensional problem. We identify dominant functional groups
representative of chemical regions in the permeability surface.  The
identification of functional groups linking to the permeability
coefficient effectively provides robust structure-property
relationships for drug-membrane permeation, and the means to perform
inverse molecular design.  Finally, we show how the apparent bias of
synthetic databases toward carboxylic groups can have deleterious
effects on the accessible range of permeability coefficients, and thus
on bioavailability.  All in all, our HTCG approach offers a
complementary approach to in vitro high-throughput screening,
providing much larger numbers of compounds ($510,000$ in this study)
than currently available in public databases.  The much larger dataset
size will help statistical models (e.g., QSPR) reach improved
transferability.  In analogy to rapidly-growing interests in
generating in silico databases of electronic
properties,\cite{faber2016machine, bartok2017machine} we expect HTCG
to have a broad impact in efficiently mapping the relevant
low-dimensional surfaces that link chemical structure to thermodynamic
properties.

\section{Methods}
\label{sec:meth}

\subsection{Molecular dynamics simulations}
\label{sec:meth_mold}
Molecular dynamics simulations in this work were performed in {\sc
  Gromacs 4.6.6}\cite{GROMACS-2008} and with the Martini force field,
\cite{M-2004, M-2007, marrink2013perspective} relying on the standard
simulation parameters.\cite{de2016martini} The integration time step
was $\delta t=0.02~\tau$, where $\tau$ is the model's natural unit of
time dictated by the units of energy $\mathcal{E}$, mass $\mathcal{M}$
and length $\mathcal{L}$, $\tau =
\mathcal{L}\sqrt{\mathcal{M}/\mathcal{E}}$. Sampling from the $NPT$
ensemble at $P=1~\text{bar}$ and $T=300~K$ was obtained by means of a
Parrinello-Rahman barostat \cite{parrinello1981polymorphic} and a
stochastic velocity rescaling thermostat,\cite{bussi2007canonical}
with coupling constants $\tau_P=12~\tau$ and $\tau_T=\tau$
respectively.  We relied on the {\sc Insane} building
tool\cite{wassenaar2015computational} to generate a membrane of
$\approx36~\text{nm}^2$ containing $N=128$ DOPC lipids (64 per layer),
$N'=1890$ water molecules, $N''=190$ antifreeze
particles,\cite{M-2007} and enough counterions to neutralize the
box. The system was subsequently minimized, heated up, and
equilibrated.
 
The potential of mean force $G(z)$ of each compound was determined by
means of umbrella sampling.\cite{torrie1977nonphysical} We employed 24
simulation windows with harmonic biasing potentials
($k=240~\text{kcal/mol/nm}^2$) centered every 0.1 nm along the normal
to the bilayer midplane. In each of them, two solute molecules were
placed in the membrane in order to increase sampling and alleviate
leaflet-area asymmetry.\cite{maccallum2008distribution,
  bereau2014more} The total production time for each umbrella
simulation was $1.2\cdot10^5~\tau$.  We then estimated the free-energy
profiles by means of the weighted histogram analysis
method.\cite{kumar1992weighted, bereau2009optimized, hub2010g_wham}
 
\subsection{Permeability coefficients} 
\label{sec:meth_perm}
The permeability coefficient is obtained from the potential of mean
force $G(z)$ and local diffusivity $D(z)$ in the resistivity
$R(z)=\exp[\beta G(z)]/D(z)$, see Eq.~\ref{eq:perm}.  For compounds
with multiple protonation states, both neutral and charged species
contribute to the total flux, leading to the total resistivity
$R_{\text{T}}$ given by\cite{carpenter2014method}
$R_{\text{T}}(z)^{-1} = R_{\text{N}}(z)^{-1} + R_{\text{C}}(z)^{-1}$,
where $R_{\text{N}}$ and $R_{\text{C}}$ are the resistivities of the
neutral and charged species, respectively. In calculating these
quantities in the case of a single (de)protonation reaction,
one has to offset the corresponding PMFs
$G_{\text{N}}(z)$ and $G_{\text{C}}(z)$ by the free-energy
difference for the acid/base reaction in bulk 
water\cite{maccallum2008distribution}
\begin{equation}
G_{\text{base}}=G_{\text{acid}}+k_{\rm B}T(\text{\pka}-\text{pH})\ln
10,
\end{equation}
see Fig.~\ref{fig:intro}, where we systematically consider neutral
$\text{pH} = 7.4$.  Beyond the distinction between acid and base, we
consider both neutral and charged species (Fig.~\ref{fig:intro}):
($i$) a neutral acid deprotonates into a charged conjugate base
(acidic \pka or \apka) and ($ii$) a neutral base protonates into a
charged conjugate acid (basic \pka or \bpka).  The extension to
zwitterions, in which two consecutive protonation and deprotonation
reactions occur in different chemical groups leaving the molecule
globally neutral, is discussed in the SI.

Estimation of the local diffusivity, $D(z)$, using the CG simulations
is a priori problematic given the tendency of these models to
inconsistently accelerate the
dynamics.\cite{rudzinski2016communication} On the other hand,
atomistic simulations showed that the diffusivity across a DOPC
bilayer was virtually independent of the chemistry of the
solute.\cite{carpenter2014method} We used this profile in the present
calculations.  We stress that the local diffusivity only provides a
logarithmic correction to $\log_{10} P$ (see Eq.~\ref{eq:perm}), and
therefore has limited impact---a variation well within $1 \log_{10}$
unit depending on the diffusivity profile.  More details can be found
in Secs.~S2 and S6 of the SI.

\subsection{Permeability surfaces}
\label{sec:meth_surf} 

We obtained the permeability surfaces presented in
Figs.~\ref{fig:permsurf} and S4 by first determining the PMF $G(z)$
for \emph{all} possible neutral combinations of one and two CG beads,
119 in total.  For each of them we then determined $G(z)$ for its
charged counterparts, amounting to a total of 232 additional
compounds.  All PMF calculations required less than $10^5$ CPU hours,
on par with the typical computational time needed to run a
\emph{single} compound at an atomistic resolution.\cite{swift2013back}
At the CG level, protonating (deprotonating) a neutral chemical group
amounts to replacing the bead type with a positive (negative)
charge. We assume that the (de)protonation reaction always occurs in
the chemical fragment represented by the more polar bead, and select
the bead accordingly.  In Sec.~S3 of the SI, we justify this approach
by analyzing the \pka distribution for various CG bead types.  By
combining neutral and charged PMFs, we calculated the permeability
coefficient of every compound as a function of the \apka (\bpka) every
$0.2$~\pka unit, and projected the results on the
$(\dGwm,\text{p}K_{\rm a})$ plane.  The data consisted of a discrete
set of permeabilities densely covering the partitioning free-energy
axis located at the $\dGwm$ of each CG compound, and were finally
interpolated on a grid with gaussian weights resulting in the surfaces
shown in Fig.~\ref{fig:permsurf} and Fig.~S4.

\subsection{Chemical space coverage}

Prediction of the water/octanol partitioning on both chemical
databases considered in this work, GDB\cite{Fink2005,Fink2007} and
ChEMBL,\cite{bento2014chembl} was performed by means of the neural
network {\sc Alogps}.\cite{Tetko2002} \apka and \bpka predictions of
neutral compounds were provided by the Calculator Plugin of {\sc
  Chemaxon Marvin}.\cite{marvin} The mean absolute error associated
with the two prediction algorithms are $0.36$~kcal/mol\cite{Tetko2002}
and $0.86$~units,\cite{liao2009comparison} respectively.  The
aggregate predictions of water/octanol partitioning and \pka
on both databases required roughly $10^2$~CPU hours.
Functional groups were identifed using the {\sc checkmol}
package.\cite{Haider2010} 511,427 molecules were coarse-grained using
the {\sc Auto-Martini} scheme.\cite{Bereau2015} {\sc Auto-Martini}
automatically determines the coarse-grained force field in two steps:
($i$) the CG mapping is optimized according to Martini-based heuristic
rules and ($ii$) interactions are set by determining a type for each
bead, selected from chemical properties of the encapsulated atoms,
especially water/octanol partitioning, net charge, and
hydrogen-bonding.

\section*{Acknowledgements}
The authors thank Clemens Rauer, Kurt Kremer, and Omar Valsson for
critical reading of the manuscript.  The authors acknowledge Chemaxon
for providing them with an academic research license for the Marvin
Suite.  This work was supported by the Emmy Noether program of the
Deutsche Forschungsgemeinschaft (DFG).  The authors gratefully
acknowledge the computing time granted by the John von Neumann
Institute for Computing (NIC) and provided on the supercomputer JURECA
at J\"ulich Supercomputing Centre (JSC).

\section*{Supporting Information}
In the supporting information (SI) we describe ($i$) our modeling of
the inhomogeneous diffusivity, $D(z)$, and provide a sensitivity
analysis; ($ii$) the representation of a (de)protonation reaction at
the coarse-grained level; ($iii$) separate permeability surfaces for
unimers and dimers; ($iv$) an extension to zwitterionic compounds;
($v$) validation of the CG simulations against both atomistic
simulations and experiments ($vi$) a validation of the predictions
extracted from the permeability surface against coarse-grained and
atomistic simulations; and ($vii$) detailed statistics of the functional
group populations corresponding to blue regions in
fig.~\ref{fig:fgroup}.

\bibliography{biblio} 

\end{document}


\title{Supporting Information for ``Drug-membrane permeability across
  chemical space''}

\author{Roberto Menichetti}
\author{Kiran H.~Kanekal}
\author{Tristan Bereau}
\email{bereau@mpip-mainz.mpg.de}
\affiliation{Max Planck Institute for Polymer Research, 55128 Mainz,
  Germany}

\date{\today}

\maketitle

\tableofcontents

\section{Introduction}
In this Supporting Information we provide additional results
integrating those presented in the main text. In Sec.~\ref{sec:diff}
we discuss the inhomogeneous diffusivity profile $D(z)$ employed in
our calculations and test the impact of different choices for this
quantity on permeability coefficients.  In Sec.~\ref{sec:prot} we
analyze our choice for representing at the CG level the (de)protonated
form of a compound.  In Sec.~\ref{sec:per_surf} we focus on the
dependence of the permeability surfaces on molecular weight and
analyze results for one- and two-beads CG compounds (``unimers'' and
``dimers'').  In Sec.~\ref{sec:zwit} we extend the discussion about
permeability surface and chemical coverage to zwitterionic compounds.
In Sec.~\ref{sec:correlation-at} we compare the permeability
coefficient $P$ of several small molecules obtained via CG molecular
dynamics simulations with independent atomistic simulation results.
In Sec.~\ref{sec:surf_vs_other} we compare the \emph{simulation-free}
predictions of permeability coefficients obtained through the permeability surface
with coarse-grained and atomistic simulation results.
In Sec.~\ref{sec:correlation-exp} we compare CG permeability
coefficients $P$ of several small molecules with independent
experimental measurements of their blood-brain-barrier permeability
coefficient ($\log_{10}\text{BB}$). In Sec.~\ref{sec:fgroup-details},
we provide detailed statistics about the dominant functional groups
present in the highlighted regions of Fig.~3 of the main text.\\

\section{Diffusivity profile}
\label{sec:diff}

\begin{figure*}[htbp]
  \begin{center}
    \includegraphics[width=0.6\linewidth]{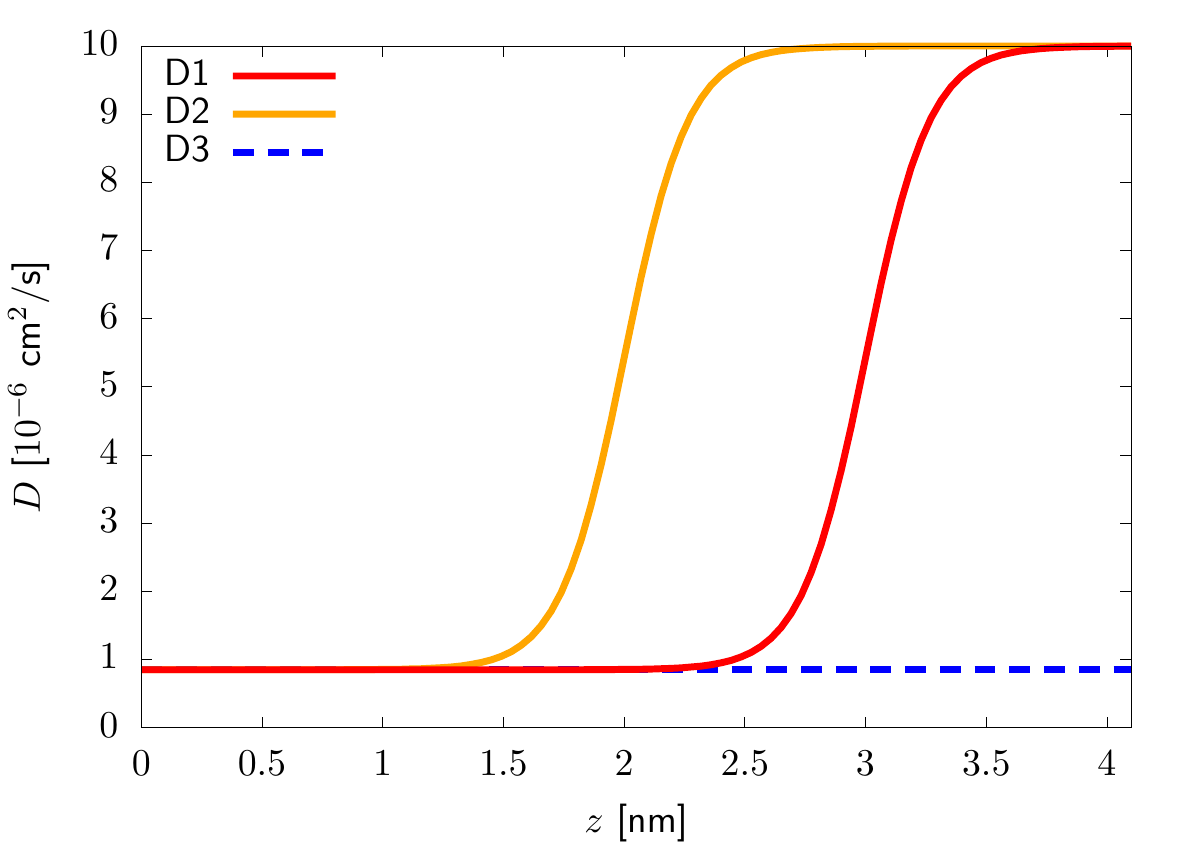}
    \caption{Local diffusivity employed in this work as a function of the normal distance $z$ of
    the compound from the bilayer midplane (D1, red line), together with the alternative 
    profiles introduced to test the sensitivity of permeability coefficients 
    to a change in $D(z)$ (see Fig.~\ref{fig:sens_diff}): a horizontally shifted version of the final parametrization (D2, orange line) and a uniform profile (D3, blue dashed line).}
    \label{fig:diff}
  \end{center}
\end{figure*}

The permeability coefficient $P$ of a small molecule in a phospholipid bilayer
is determined from its potential of mean force $G(z)$ 
and inhomogeneous diffusivity $D(z)$ as \cite{votapka2016two}
\begin{equation}
  \label{eq:perm}
  P^{-1} = \int {\rm d}z \frac{\exp[\beta G(z)]}{D(z)},
\end{equation}
with $\beta=1/k_BT$. For compounds with multiple protonation states, both neutral and charged
species contribute to the total flux and must be 
taken into account with a slight modification of Eq.~\ref{eq:perm} (see Methods in the main text).
As the Martini force field is explicitly parametrized to 
reproduce the thermodynamics of partitioning between solvents of different polarity
 \cite{M-2004, M-2007,marrink2013perspective}, it is
capable of providing accurate results for the potential of mean force. On the other hand, the
calculation of $D(z)$ starting from CG molecular dynamics simulations is highly nontrivial due 
to the tendency of these models to inconsistently accelerate the dynamics \cite{rudzinski2016communication}.
The determination of the inhomogeneous diffusivity of a compound thus still relies on performing expensive atomistic molecular dynamics 
simulations. However, by comparing the results for several small molecules embedded in a DOPC bilayer it was recently shown that $D(z)$ is virtually insensitive to the chemical detail \cite{carpenter2014method}.
Moreover, from Eq.~\ref{eq:perm} it is apparent that $D(z)$ only provides a linear correction to the permeability coefficient,
therefore a logarithmic correction to the corresponding order of magnitude ($\log_{10} P$).
Starting from these considerations, in the calculation of permeability coefficients we 
considered a unique effective diffusivity profile across chemical space (and for different protonation states).
We parametrized it as
\begin{equation}
D(z)=\alpha+\frac{\beta}{e^{-\gamma(x-\delta)}+1},
\end{equation}
which correctly captures the main features of its atomistic
counterpart, \emph{i.e.}, an increased diffusivity of the compound as
it leaves the lipid tails and enters the bulk water environment. We
tuned the parameters $\alpha,~\beta,~\gamma$ and $\delta$ to
reproduce the atomistic simulation results of
Ref.~\citenum{carpenter2014method}. The final, optimized form of
$D(z)$ employed in our calculations is shown in
Fig.~\ref{fig:diff}, with $\alpha=0.85\cdot10^{-6}~\text{cm}^2/s$, $\beta=9.15\cdot10^{-6}~\text{cm}^2/s$, $\gamma=7.5~\text{nm}^{-1}$, $\delta=3~\text{nm}$.

\begin{figure*}[htbp]
  \begin{center}
    \includegraphics[width=0.6\linewidth]{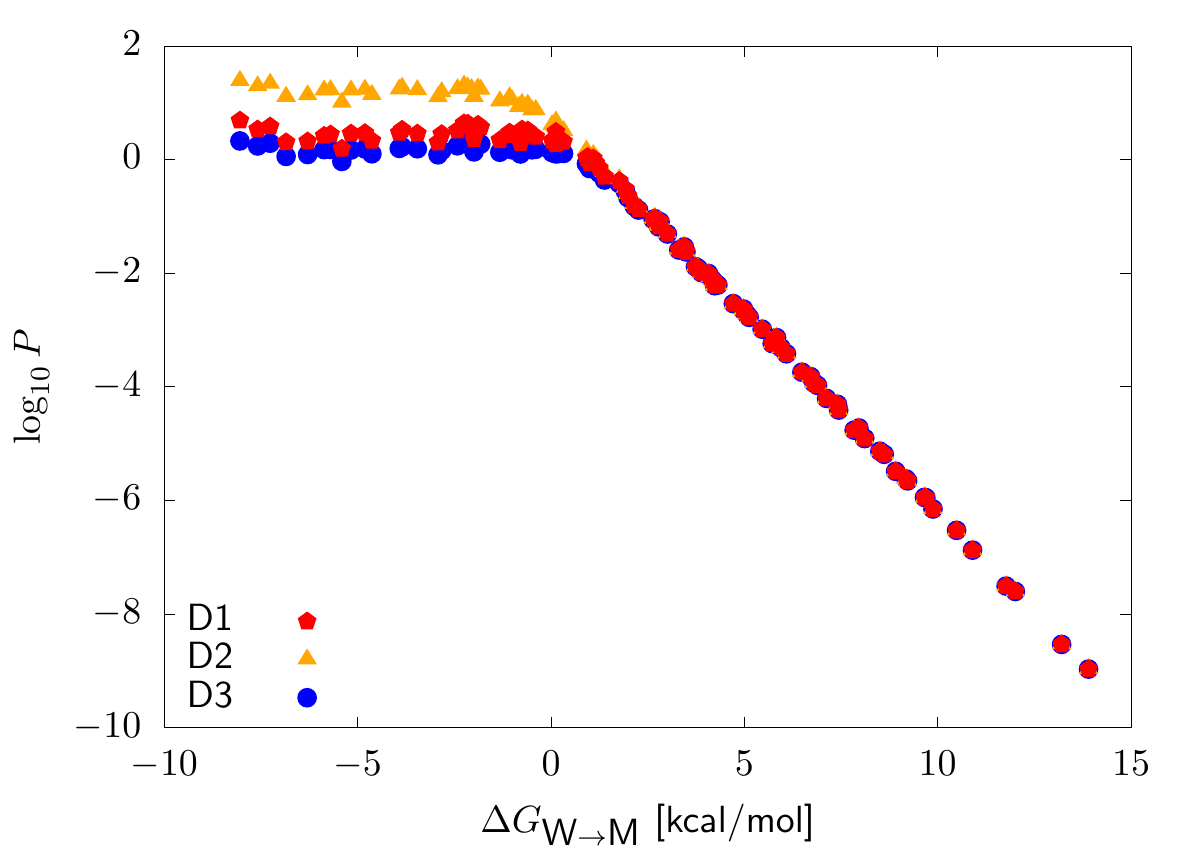}
    \caption{Permeability coefficient ($\log_{10}$ scale) of a neutral
      compound as a function of its water/membrane free energy $\dGwm$
      obtained by considering the three different diffusivities $D(z)$
      shown in Fig.~\ref{fig:diff}: the optimized parametrization
      (D1), its horizontally shifted version (D2), and a homogeneous
      profile (D3). Key labels and colors coding follow those of
      Fig.~\ref{fig:diff}.}
    \label{fig:sens_diff}
  \end{center}
\end{figure*}

We further tested the impact of this quantity
on the permeability coefficients---and consequently on the
permeability surfaces---by considering two alternative profiles: an
horizontally shifted version of the optimized parametrization, and a
homogeneous profile. Both curves are shown in
Fig.~\ref{fig:diff}. The permeabilities obtained by relying on the
three different diffusivities in the case of neutral compounds are
presented in Fig.~\ref{fig:sens_diff} as a function of the
water/membrane partitioning free energy $\dGwm$.
\label{sec:test_diff}
It is apparent that a change in $D(z)$ only affects the permeability of hydrophobic compounds ($\dGwm\lesssim0$). Most importantly, deviations are within one $\log_{10}$ unit, which is our degree of accuracy in predicting permeability coefficients. These results are in agreement with previous studies that correlated a variation of one order of magnitude in the inhomogeneous diffusivity to roughly one $\log_{10}$ unit in the permeability coefficient \cite{lee2016simulation}.
 
\section{Representing (de)protonation at the CG level}
\label{sec:prot}
\begin{figure*}[htbp]
  \begin{center}
    \includegraphics[width=1.0\linewidth]{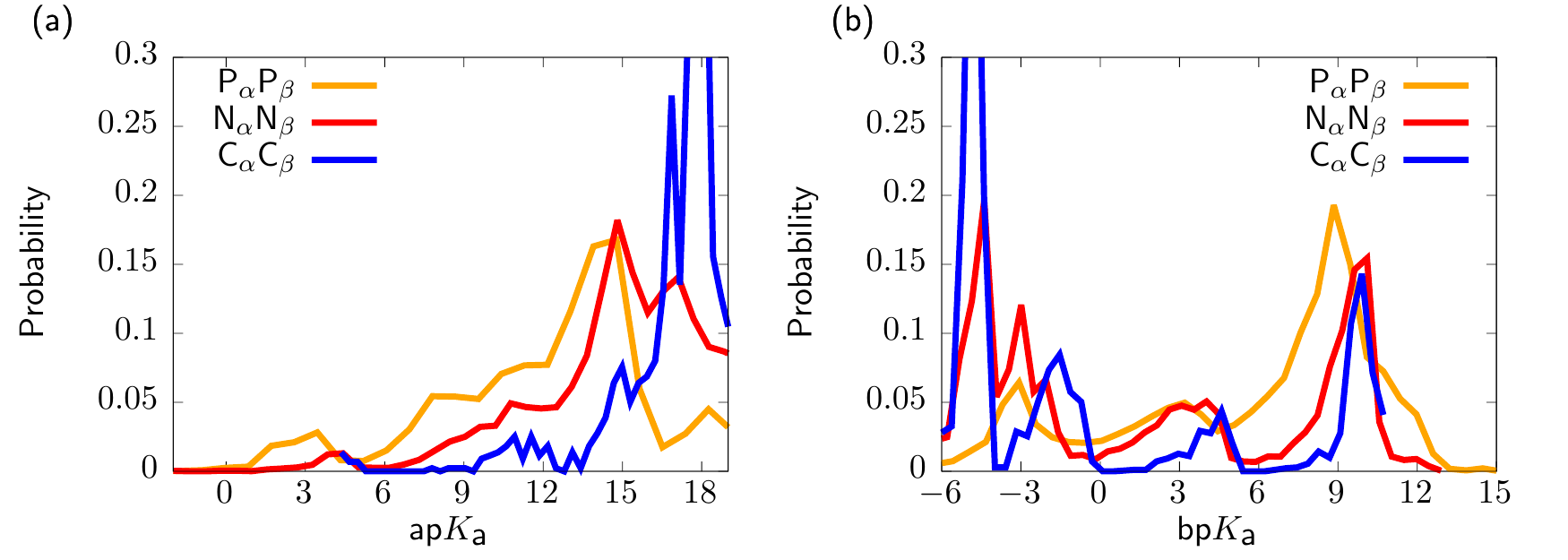}
    \caption{\apka (a) and \bpka (b) histograms for all compounds in the GDB 
    database mapping to coarse-grained dimers that only contain a combination of beads with the same degree of polarity: Polar ($P_{\alpha}P_{\beta}$),
    nonpolar ($N_{\alpha}N_{\beta}$) and apolar ($C_{\alpha}C_{\beta}$) compounds---following the standard Martini notation \cite{M-2004, M-2007,marrink2013perspective}.}
    \label{fig:prot_hyst}
  \end{center}
\end{figure*}
The Martini mapping of a small molecule is obtained by decomposing it
in chemical fragments and associating a bead type appropriately
selected from the Martini list to each fragment \cite{Bereau2015}. As
in the calculation of the permeability coefficient one has to account
for the different protonation states of a compound (see Methods
section in the main text), it is necessary to discuss how to represent
(de)protonation reactions at the coarse-grained level: in other words,
how to determine the Martini parametrization of the (de)protonated
form of a compound starting from the neutral case. In the following,
we will focus on the set of coarse-grained molecules considered in
this work, \emph{i.e.}, unimers and dimers.

Within a neutral
compound, (de)protonating a chemical group leaves the corresponding
chemical fragment with a (negative) positive charge. Therefore, at the
coarse-grained level the neutral bead encapsulating the fragment has
to be accordingly replaced with a charged one.  In Martini, the
charged bead can be selected among four different types, mimicking
different hydrogen-bond capabilities: $Q_0$ (no hydrogen bond), $Q_{\text{da}}$
(donor/acceptor), $Q_{\text{a}}$ (acceptor) or $Q_{\text{d}}$
(donor) \cite{M-2004, M-2007,marrink2013perspective}.\\ We
systematically represent the protonated form of a chemical fragment by
means of a positively charged donor bead type ($Q_{\text{d},+1}$),
while the deprotonated form with a negatively charged acceptor bead
type ($Q_{\text{a},-1}$).  Although (de)protonation could allow the
fragment to have both hydrogen-bond donor \emph{and} acceptor
capabilities, employing a $Q_{\text{da}}$ bead type does not
significantly impact the resulting permeability coefficients.\\ While
for small molecules that map to unimers the charged bead type uniquely
determines the coarse-grained representation of their (de)protonated
form, in the case of dimers there is an additional degree of freedom
associated to the choice of the site.  Indeed, starting from a neutral
dimer $B_1B_2$ there are two different possibilities depending on
which chemical fragment is subject to (de)protonation: $Q_1B_2$ and
$B_1Q_2$. Rather than considering all possible cross-combinations of
neutral and charged bead types, in the case of dimers we employ a
data-driven approach to estimate the bead type most likely to
(de)protonate.  We thus extracted from the GDB database all compounds
mapping to polar (combination of $P_{\alpha}P_{\beta}$ bead types,
following the standard Martini notation \cite{M-2004,
  M-2007,marrink2013perspective}) nonpolar ($N_{\alpha}N_{\beta}$) and
apolar ($C_{\alpha}C_{\beta}$) dimers, and separately calculated \apka
and \bpka histograms in the three cases. The resulting distributions
shown in Fig.~\ref{fig:prot_hyst} suggest a correlation between the
polarity of the Martini bead type and the acid/base strength of the
encapsulated chemical fragment. Strongest acids/bases tend to be
represented by polar beads $P_{\alpha}$, followed by non-polar
$N_{\alpha}$ and apolar $C_{\alpha}$ ones. Starting from these
results, in the case of dimers we assumed that (de)protonation always
occurs within the chemical fragment represented by the more polar bead
type.

We relied on this set of assumptions for the calculation of
permeability surfaces (Fig.~\ref{fig:surf_comb} and Fig.~2 in the main
text) and in the databases of permeability coefficients for the
510,000 small molecules extracted from the GDB database.

\section{Permeability surfaces: unimers and dimers}
\label{sec:per_surf}

\begin{figure*}[htbp]
  \begin{center}
    \includegraphics[width=0.8\linewidth]{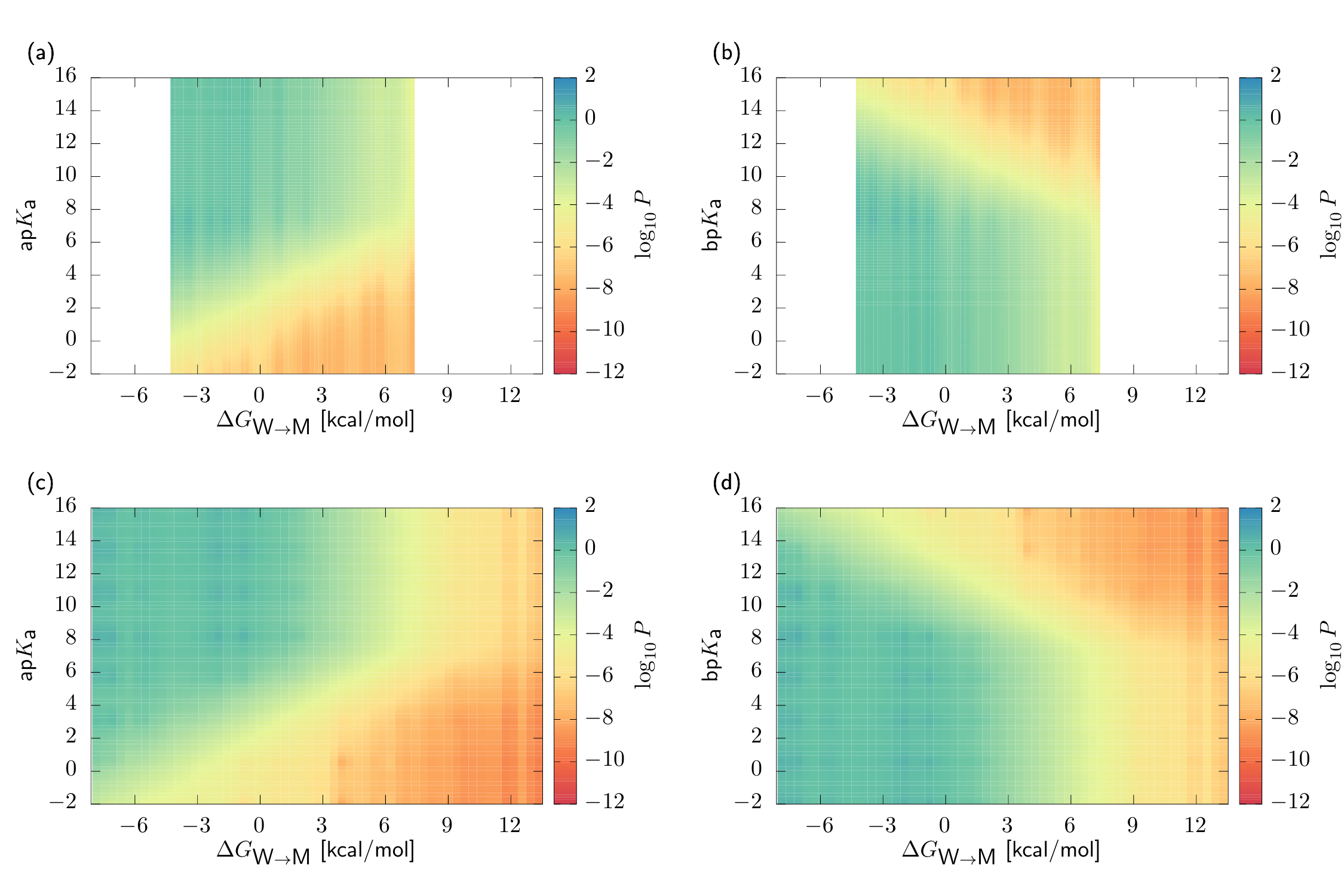}
    \caption{Permeability surfaces ($\log_{10}$ scale) for Martini
      unimers (a,b) and dimers (c,d) as a function of two
      small-molecule descriptors: the acidic or basic \pka in water
      and the water/membrane partitioning free energy. We employ a
      common range in the horizontal axis for unimers and dimers in
      order to underline the transferability of the surfaces across
      molecular weights (see text).  Cooler (warmer) colors correspond
      to faster (slower) permeating molecules.}
    \label{fig:surf_comb}
  \end{center}
\end{figure*}

We discuss how the permeability surfaces (main text, Fig.~2) depend on
molecular weight. At the CG level, this corresponds to an increase in
the number of beads of the small molecule.  For both unimers to
dimers, Fig.~\ref{fig:surf_comb} displays permeability surfaces as a
function of the acidic or basic \pka of the compound in water (\apka
and \bpka) and its water/membrane partitioning free energy
$\dGwm$.

In analogy with what we observe in the case of chemical
coverage (see Fig.~3 in the main text), going from unimers to dimers
broadens the range of water/membrane partitioning free-energies. This
stems from the extensive nature of the partitioning coefficient.
On the other hand, a comparison of the surfaces in the range
$-4\lesssim\dGwm\lesssim7$\ [kcal/mol]---separately for panels (a,c)
and (b,d)---highlights the extremely good transferability of the
results across molecular weights, as the profiles corresponding to
unimers and dimers superimpose to a large extent.

\section{Zwitterionic compounds}
\label{sec:zwit}

\begin{figure}[htbp]
  \begin{center}
    \includegraphics[width=0.5\linewidth]{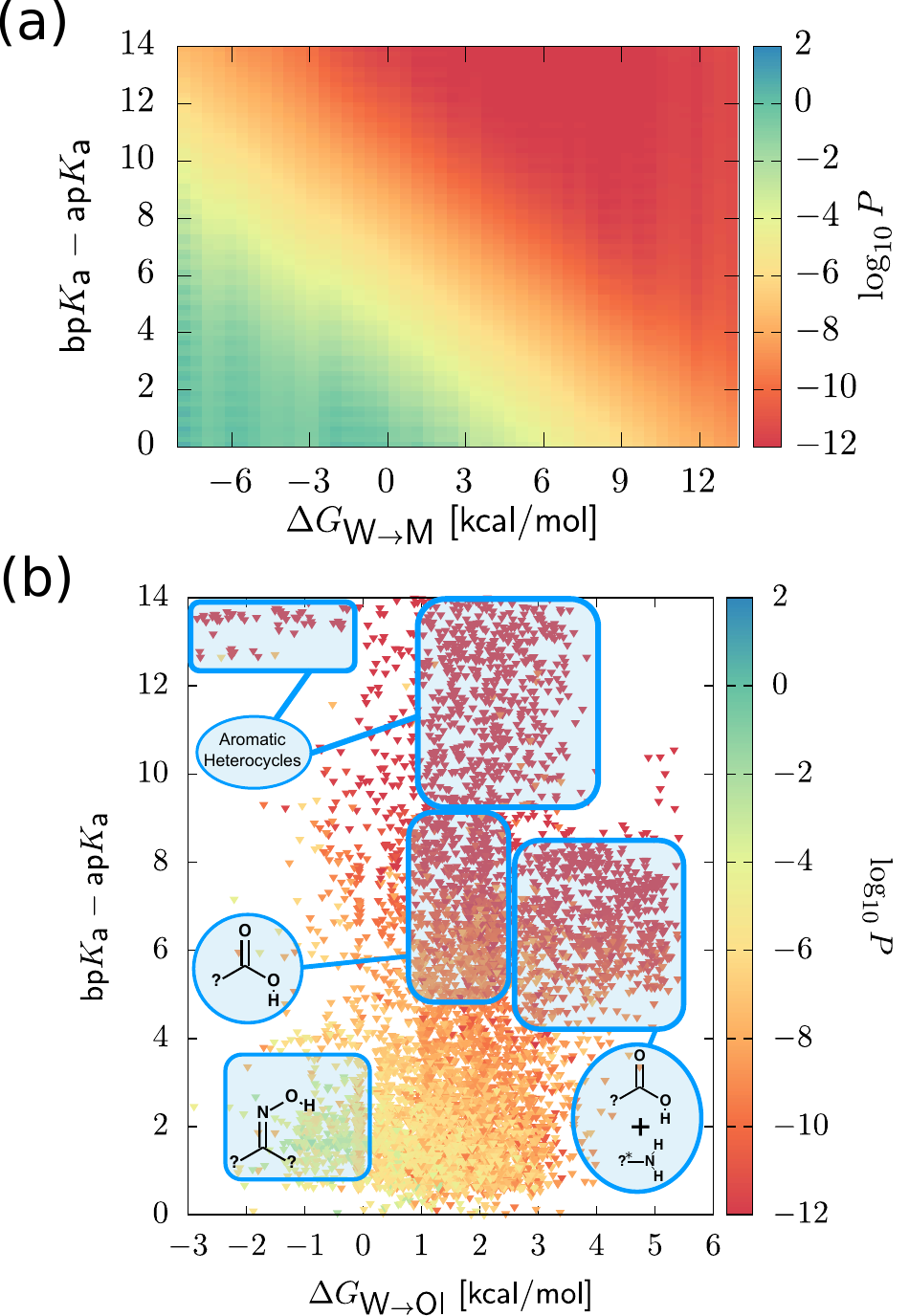}
    \caption{(a) Permeability surface of zwitterionic compounds. Captions follow of those of
      Fig.~\ref{fig:surf_comb}. (b)  Chemical-space coverage of GDB  zwitterions projected onto \pka and
      water/octanol partitioning free energies $\dGwol$. Regions
      highlighted in light blue display several representative
      chemical groups.  Substitutions denoted by ``{\bf ?}''
      correspond to H, alkyl, or aryl groups, while ``{\bf ?*}'' only
      correspond to alkyl groups.}
    \label{fig:zwit}
  \end{center}
\end{figure}

We now extend the discussion about permeability surface and chemical coverage (Figs.~2,3 in the main text) to zwitterionic compounds,
which consist of two acid/base reactions while keeping the molecule electrically neutral.

In analogy to the case of compounds presenting a single (de)protonation reaction, 
both the neutral and doubly-charged species
contribute to the permeability coefficient through the total resistivity (see Eq.~2 in the main text). The corresponding potentials
of mean force $G_{\text{N}}(z)$ and $G_{\text{C}}(z)$ must be offset by the free-energy difference between the
neutral compound and the zwitterionic form in bulk water.
By assuming independence of the individual reactions, this offset can be written as
\begin{equation}
\label{eq:zwit}
G_{\text{neut}}=G_{\text{zwit}}+k_{\rm B}T(\text{\bpka}
-\text{\apka})\ln10,
\end{equation}
which depends on the \emph{difference} in ionization species, $\text{\bpka} - \text{\apka}$.
We require that the zwitterionic form is more stable than both the
neutral and single-charged species, leading to the additional
constraint ${\text{\bpka} \gtrsim 7.4}$ and $\text{\apka} \lesssim
7.4$.\\
As in the case of acidic and basic compounds (see Sec.~\ref{sec:prot}), it is necessary to discuss how to represent the zwitterionic form of a compound at the coarse-grained level. We again distinguish between unimers and dimers.\\
In unimers, the protonation and deprotonation reactions occur within the same chemical fragment, so that the coarse-grained representation of the doubly-charged form of a compound coincides with the original, neutral one. The minimum length scale accessible by the coarse-grained model (roughly of the size of the chemical fragment) thus precludes to represent a zwitterionic compound in terms of a single bead.\\
In the case of dimers, we assumed that the protonation and deprotonation reactions occur in chemical fragments belonging to two different beads, thus leading to a CG compound containing two opposite charges starting from the neutral one.
Together with our choice of the bead types for representing (de)protonated chemical fragments (see Sec.~\ref{sec:prot}), this means that given a neutral $B_1B_2$ dimer its zwitterionic form is $Q_{\text{a},-1}Q_{\text{d},+1}$.\\
We relied on these assumptions to calculate the permeability surface for zwitterionic  
compounds that map to coarse-grained dimers. Results are shown in Fig.~\ref{fig:zwit}a.
Akin to the single-protonation compounds (Fig.~\ref{fig:surf_comb}), the surface shows a strong
dependence on both descriptors ($\text{\bpka}-\text{\apka}$ and
$\dGwm$) with hydrophobic compounds generally leading to larger
permeability coefficients.  These also require smaller values of
$\text{\bpka} - \text{\apka}$, i.e., weaker acids and bases.  
Stronger acid/base compounds significantly lower the permeability coefficient,
reaching values often lower than for the single-protonation cases
(Fig.~\ref{fig:surf_comb}).\\
The corresponding chemical-space coverage is shown in Fig.~\ref{fig:zwit}b.
Unsurprisingly, it displays many functional groups that are present in
\emph{both} \apka and \bpka coverages (Fig.~3 in the main text).
Notable examples are carboxylic groups as well as certain aromatic
heterocyclic compounds.

\section{Comparison of coarse-grained and atomistic simulation results}
\label{sec:correlation-at}

\begin{figure*}[htbp]
  \begin{center}
    \includegraphics[width=0.65\linewidth]{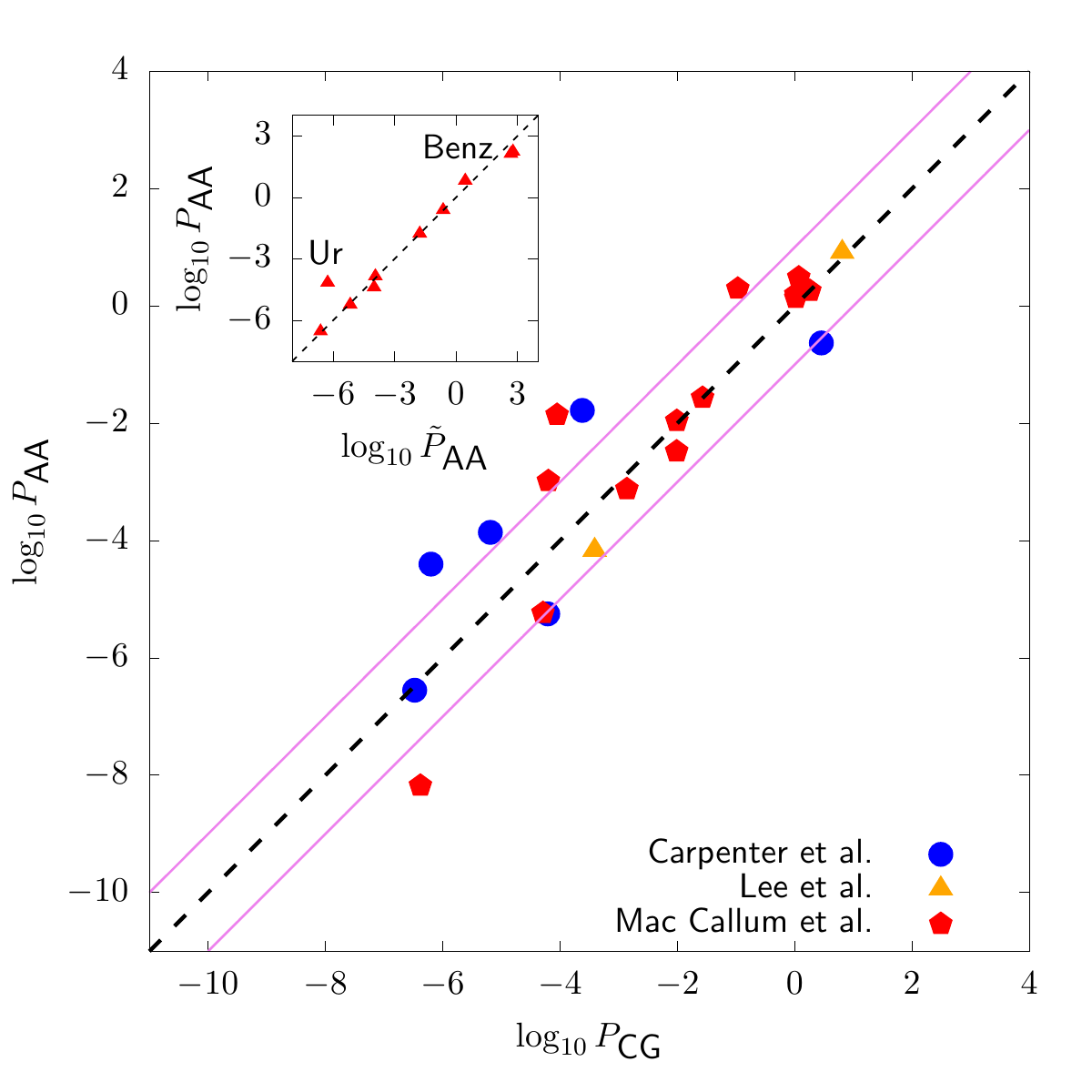}
    \caption{Inset: Correlation between permeability coefficients $\log_{10} \tilde{P}_{\text{AA}}$ calculated from AA simulations \cite{carpenter2014method,lee2016simulation}, and the $\log_{10} P_{\text{AA}}$ obtained by combining AA potentials of mean force $G(z)$ with the effective diffusivity profile $D(z)$ presented in Fig.~\ref{fig:diff}. Urea and Benzoic acid---see text---are marked with the labels ``Ur'', ``Benz''. Main: Correlation between permeability coefficient calculated via AA and CG potentials of mean force ($\log_{10} P_{\text{AA}}$ and $\log_{10} P_{\text{CG}}$, respectively), in both cases relying on the effective diffusivity profile presented in Fig.~~\ref{fig:diff}. We present results for the compound extracted from Ref.~\citenum{carpenter2014method} (``Carpenter et al.''), Ref.~\citenum{lee2016simulation} (``Lee et al.'') and Ref.~\citenum{maccallum2008distribution} (``Mac Callum et al.'').} 
    \label{fig:corr_plot_aa_cg}
  \end{center}
\end{figure*}

In this work, we determined permeability coefficients through Eq.~\ref{eq:perm} by considering a unique effective diffusivity profile across chemical space (see Sec.~\ref{sec:diff}), combining it with potentials of mean force $G(z)$ extracted from CG simulations.
In order to assess the accuracy of these results, we now test them against independent atomistic simulation ones \cite{carpenter2014method,lee2016simulation}.

Having employed a unique diffusivity profile in our calculations, it is necessary to first discuss the sensitivity of atomistic permeability coefficients with respect to a replacement of the original diffusivity $D(z)$ of a compound with our effective profile. 
We performed this analysis on a subset of the compounds that in Ref.~\citenum{carpenter2014method,lee2016simulation} were investigated by means of atomistic simulations, excluding from the calculations all molecules containing multiple intertwined rings due to difficulties in obtaining their coarse-grained representation. It is important to stress that the compounds investigated in Ref.~\citenum{lee2016simulation} were embedded in a dimyristoylphosphatidylcholine (DMPC) bilayer, a different lipid with respect to the one considered in this work (DOPC). In the case of Urea and Benzoic (``Ur'' and ``Benz'' in the inset of Fig.~\ref{fig:corr_plot_aa_cg}), this required us to slightly modify the effective diffusivity profile to account for the difference in membrane composition.

For the set of compounds considered, in the inset of
Fig.~\ref{fig:corr_plot_aa_cg} and in Table \ref{table:compare} we compare the atomistic permeability
coefficients $\log_{10}\tilde P_{\text{AA}}$ reported in
Ref.~\citenum{carpenter2014method,lee2016simulation} with the
$\log_{10} P_{\text{AA}}$ we obtained by means of our effective
diffusivity---\emph{i.e.}, calculated by means of the
atomistic $G(z)$ and the profile presented in
Fig.~\ref{fig:diff}.  The excellent correlation between these two
quantities confirms that the introduction of the effective diffusivity
doesn't significantly impact the permeability coefficient, which
largely depend on the potential of mean force $G(z)$. In the case of
Urea \cite{lee2016simulation} (``Ur'' in
Fig.~\ref{fig:corr_plot_aa_cg}), the only statistically significant
outlier, the observed discrepancy is due to difficulties to visually
extract the atomistic $D(z)$ close to the bilayer midplane $z\approx
0$ in Ref.~\citenum{lee2016simulation}.

We now compare atomistic and coarse-grained simulation predictions for
permeability coefficients to assess the accuracy of the
latter. As atomistic reference data, we first considered the
previously introduced set of compounds extracted from
Refs.~\citenum{carpenter2014method, lee2016simulation}. We
systematically coarse-grained all these compounds through the {\sc
  Auto-Martini} tool \cite{Bereau2015}, again excluding from the
calculations all chemical compounds containing multiple intertwined
rings.
In the case of atenolol and salbutamol, we had to account for
the presence of discrepancies in the {\sc Alogps} \cite{Tetko2002}
prediction of water/octanol partitioning free energy against
experimental measurements for specific chemical fragments by slightly
fine-tuning the {\sc Auto-Martini} output. For completeness, in
Sec.~\ref{sec:gromacs} we report {\sc Gromacs} input files with the
final force-field parametrization for the entire set of small
molecules. Subsequently, we performed CG molecular dynamics
simulations as described in the Methods section of the main text and
calculated the corresponding CG permeability coefficients $\log_{10}
P_{\text{CG}}$. 

Given that the set contains only a limited number of
compounds---most of them being beyond the upper limit in molecular
weight considered in this work---we further included in the analysis
the subset of amino-acid side chains discussed in
Ref.~\citenum{menichetti2017silico}, whose behavior in a DOPC membrane
was analyzed in Ref.~\citenum{maccallum2008distribution} by means of
atomistic simulations. Unfortunately,
Ref.~\citenum{maccallum2008distribution} doesn't provide results for
the atomistic diffusivity $D(z)$. However, having established that the
use of the effective diffusivity provides consistent permeability
coefficients within the degree of accuracy pursued in this work (inset
of Fig.~\ref{fig:corr_plot_aa_cg}), we employed this profile together
with the atomistic $G(z)$ to determine the permeability coefficients
$\log_{10} P_{\text{AA}}$ of amino-acid side chains. The corresponding
coarse-grained $\log_{10} P_{\text{CG}}$ were again determined by
means of CG simulations.

A comparison between permeability coefficients obtained by means of
atomistic and coarse-grained simulations ($\log_{10} P_{\text{AA}}$
and $\log_{10} P_{\text{CG}}$, respectively) for all 21 compounds is
presented in Fig.~\ref{fig:corr_plot_aa_cg} and Table~\ref{table:compare}. For consistency among
atomistic results, the $\log_{10} P_{\text{AA}}$ of all compounds are
calculated by considering the atomistic potential of mean force
$G(z)$, together with the effective diffusivity profile.
Overall, Fig.~\ref{fig:corr_plot_aa_cg} suggests a high correlation between atomistic and CG results, which extends over a wide range of orders of magnitude and beyond the range of molecular weights investigated in this work---Pearson correlation coefficient $R^2 \approx 0.9$, with a mean absolute error of roughly one $\log_{10}$ unit---thus confirming the accuracy of CG models in predicting the permeability coefficient of a compound. 

\section{Permeability surface against coarse-grained/atomistic simulations}
\label{sec:surf_vs_other}
The surfaces presented in Fig.~S4, S5 and Fig.~2 of the main text enable the calculation of the permeability coefficient of a compound only starting from two key molecular properties: the water/octanol partitioning free energy $\dGwol$ and acid dissociation constant \pka.
We now validate these \emph{simulation-free} predictions against the ones obtained by performing explicit simulations, either atomistic or coarse-grained, again considering the 21 chemical compounds introduced in Sec.~\ref{sec:correlation-at}.

As the permeability surfaces were derived for all compounds ranging from 30 to 160 Da that map to coarse-grained unimers and dimers, we accordingly distinguish between the small molecules that satisfy both these constraints (set M1) and the ones who don't (set M2). While M1 will allow for a direct validation of the permeabilities obtained from the surface, M2 will give insights into the transferability of our results across molecular weight and coarse-grained representations.
\begin{figure*}[htbp]
  \begin{center}
    \includegraphics[width=0.9\linewidth]{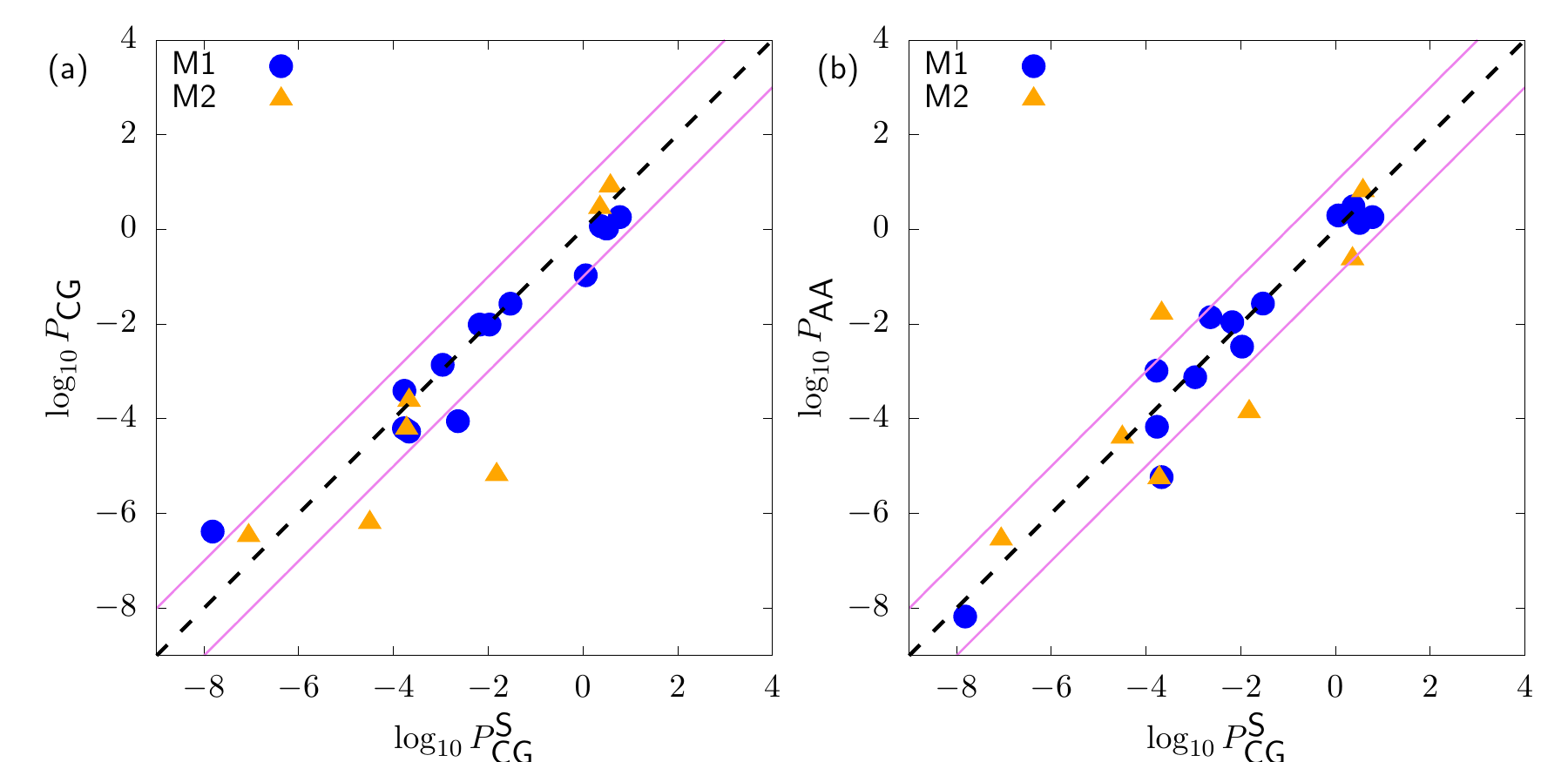}
    \caption{Comparison of the \emph{simulation-free} permeability coefficients $\log_{10} P^{\text{S}}_{\text{CG}}$ extrapolated from the permeability surface with coarse-grained $\log_{10} P_{\text{CG}}$ (a) and atomistic $\log_{10} P_{\text{AA}}$ (b) simulation results, both calculated relying on the effective diffusivity profile. We present results for small molecules extracted from Refs.~\citenum{carpenter2014method,lee2016simulation,maccallum2008distribution}, dividing them according to whether they are within (M1) or outside (M2) the range of molecular weights ($30-160$~Da) and coarse-grained representations (unimers and dimers) investigated in this work.}
    \label{fig:corr_plot_surface}
  \end{center}
\end{figure*}

For each of the 21 compounds, we first determined the molecular water/octanol partitioning free-energy and \pka by means on the {\sc Alogps} \cite{Tetko2002} and {\sc Chemaxon Marvin} \cite{marvin} prediction tools. Calculating a permeability coefficient from the surface requires to convert $\dGwol$ into the water/membrane partitioning free energy $\dGwm$: this can be done by relying on the linear relations presented in Ref.~\citenum{menichetti2017silico}. These relations are molecular-weight dependent, and in Ref.~\citenum{menichetti2017silico} were determined for all small molecules in the range $30-160$~Da that map onto coarse-grained unimers and dimers.
For the 14 compounds belonging to the set M1, this allowed us to directly calculate their $\dGwm$ starting from $\dGwol$. 

In the set M2, salicylate and benzoic acid are within the correct range of molecular weight but present a different coarse-grained representation (four beads due to the presence of an aromatic ring). Mannitol has a slightly larger molecular weight ($182$~Da), and maps onto a coarse-grained dimer. Atenolol, cimetidine, ibuprofen and salbutamol present both molecular weights higher than 160 Da and different coarse-grained representations. As the water/octanol to water/membrane relationship is unknown in these cases, we calculated their $\dGwm$ from $\dGwol$ by relying on the one appropriate for coarse-grained dimers.

A projection of the calculated $\dGwm$ and \pka of each small molecule on the corresponding surface allowed us to estimate its permeability coefficient $\log_{10} P^{\text{S}}_{\text{CG}}$. Results for all 21 compounds are presented in Table~\ref{table:compare}. 
Furthermore, in Fig.~\ref{fig:corr_plot_surface}a,b we separately compare the extrapolated $\log_{10} P^{\text{S}}_{\text{CG}}$ with the permeabilities obtained by means of coarse-grained ($\log_{10} P_{\text{CG}}$) and atomistic ($\log_{10} P_{\text{AA}}$) simulations, in both cases calculated by relying on the effective diffusivity profile. 

Fig.~\ref{fig:corr_plot_surface}a shows that for all compounds belonging to the set M1 the \emph{simulation-free} prediction of the permeability coefficient is in good agreement with the one calculated via explicit coarse-grained simulations (restricted to M1, we have $R^2\approx0.92$ with a mean absolute error of $1 \log_{10}$ unit). Therefore, an a posteriori extrapolation from the permeability surface has a negligible effect on the coarse-grained results. 

Within the set M2, molecules slightly above the range of molecular weights investigated in this work are characterized by level of accuracy that is comparable to the one of the set M1. However, further increasing the molecular weight (and coarse-grained representations) of the compound concurrently generates an increase in the deviation between extrapolated and coarse-grained permeabilities. Going from mannitol ($182$~Da) to salbutamol and cimetidine ($239$ and $252$~Da, respectively), the discrepancy between the two predictions grows from $0.4$ up to roughly $3~\log_{10}$ units.
This is a consequence of approximating the unknown $\dGwm$ to $\dGwol$ relation in the case of high molecular-weight molecules with the one appropriate to describe smaller ones. In all cases, the molecular-weight-dependent mapping from water/octanol to water/membrane thus limits the transferability of our
predictions \cite{menichetti2017silico}.

Interestingly, the $\log_{10} P^{\text{S}}_{\text{CG}}$ predictions for urea and benzoic acid extracted from the surface characteristic of the DOPC bilayer are in good agreement with the $\log_{10} P_{\text{CG}}$ obtained by directly simulating a DMPC membrane.

In Fig.~\ref{fig:corr_plot_surface}b we further compare the permeability coefficients $\log_{10} P^{\text{S}}_{\text{CG}}$ extrapolated from the surface with the corresponding $\log_{10} P^{\text{S}}_{\text{CG}}$ obtained by means of atomistic simulations. Overall, the accuracy is similar to the one associated to coarse-grained simulations (Fig.~\ref{fig:corr_plot_aa_cg})---$R^2\approx0.9$, with a mean absolute error of $1~\log_{10}$ unit.

\begin{table}[t]
\begin{center}
\begin{tabular}{ c | c c c c}
\hline\hline
 & $\log_{10} \tilde{P}_{\text{AA}}$ & $\log_{10} P_{\text{AA}}$ & $\log_{10} P_{\text{CG}}$ & $\log_{10} P^{\text{S}}_{\text{CG}}$\\
\hline
Atenolol & -1.77 & -1.78 & -3.62 & -3.67\\
Cimetidine & -3.94 & -3.86 & -5.19 & -1.82\\
Ibuprofen & -0.63 & -0.63 & 0.45 & 0.36\\
Mannitol & -6.62 & -6.55 & -6.48 & -7.06\\
Salbutamol & -4 & -4.40 & -6.20 & -4.50\\
Salicylate & -5.17 & -5.25 & -4.21 & -3.72\\
Urea & -6.27 & -4.17 & -3.41 & -3.77\\
Benzoic acid & 0.45 & 0.81 & 0.91 & 0.58\\
ser & --- & -1.96 & -2.01 & -2.18\\
cys & --- & 0.29 & -0.97 & 0.06\\
met & --- & 0.26 & 0.26 & 0.78\\
thr & --- & -1.56 & -1.57 & -1.53\\
asn & --- & -3.12 & -2.86 & -2.96\\
val & --- & 0.14 & 0.02 & 0.51\\
leu & --- & 0.20 & 0.02 & 0.49\\
ile & --- & 0.49 & 0.07 & 0.38\\
gln & --- & -2.48 & -2.01 & -1.97\\
arg & --- & -8.17 & -6.38 & -7.82\\
asp & --- & -2.98 & -4.20 & -3.78\\
glu & --- & -5.23 & -4.27 & -3.67\\
lys & --- & -1.85 & -4.05 & -2.64\\
\hline
\end{tabular}
\vspace{0.7cm}
\caption{Permeability coefficients ($\log_{10}$ scale) calculated for the validation set of 21 compounds analyzed in this work. We report atomistic simulation results \cite{carpenter2014method,lee2016simulation} ($\log_{10} \tilde{P}_{\text{AA}}$), atomistic ($\log_{10} P_{\text{AA}}$) and coarse-grained ($\log_{10} P_{\text{CG}}$) simulation results calculated by relying on the effective diffusivity profile, and \emph{simulation-free} predictions $\log_{10} P^{\text{S}}_{\text{CG}}$ extrapolated from the permeability surface.
}
\label{table:compare}
\end{center}
\end{table}

\section{Comparison with experimental results}
\label{sec:correlation-exp}

\begin{figure*}[htbp]
  \begin{center}
    \includegraphics[width=0.6\linewidth]{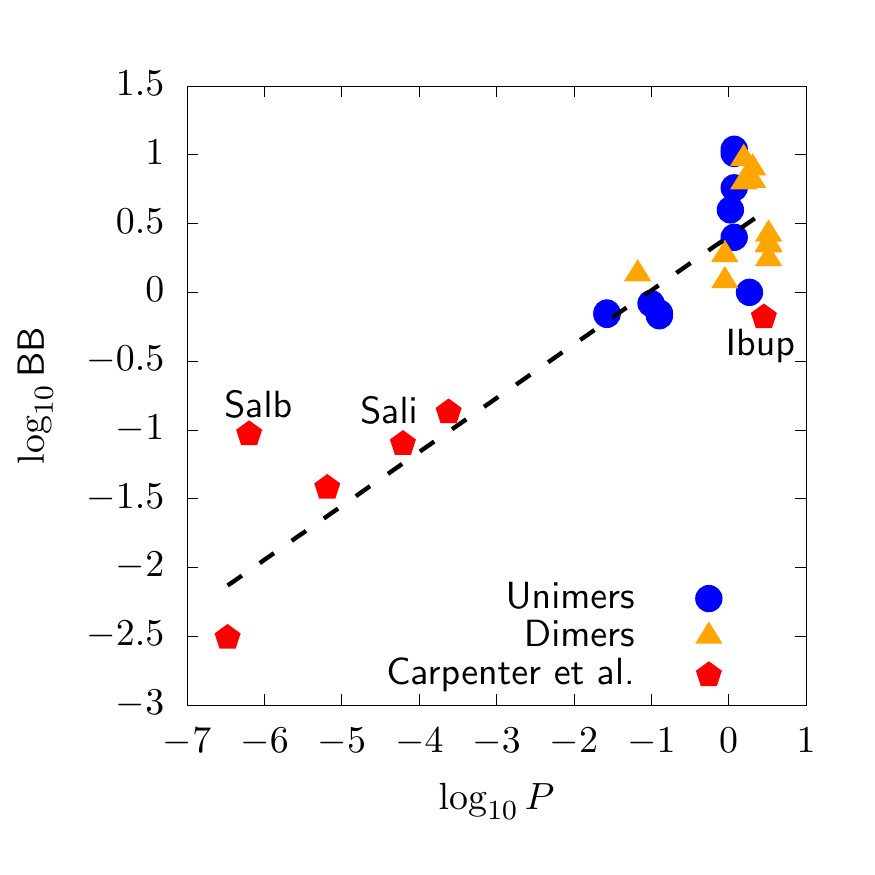}
    \caption{Correlation between permeability coefficients ($\log_{10}$ scale) calculated from CG simulations and experimental blood-brain-barrier permability coefficients $\log_{10}\text{BB}$. We present results for small molecules extracted from the database of experimental $\log_{10}\text{BB}$ reported in Ref.~\citenum{wichmann2007prediction}, dividing them according to their CG representation: unimers (blue circles) and dimers (orange triangles). We further present calculations for a subset of the compounds analyzed in Ref.~\citenum{carpenter2014method} (red pentagons).
Salicylate and ibuprofen, explicitly discussed in the main text, are marked with the labels ``Sali'' and ``Ibu''. Salbutamol, the only statistically significant outlier---see text---is marked with the label ``Salb''.}
    \label{fig:corr_plot}
  \end{center}
\end{figure*}

The permeation rates calculated for several chemically different
compounds through Eq.~\ref{eq:perm} and atomistic molecular dynamics
simulations were shown to exhibit extremely high correlations with
their experimental counterparts, \emph{e.g.}, blood-brain-barrier ($\log_{10}\text{BB}$)
or PAMPA permeability coefficients \cite{carpenter2014method,bennion2017predicting}. It is therefore
important to perform the same analysis on the results we obtain by
employing CG models.

We first considered the dataset of experimental
$\log_{10}\text{BB}$ permeabilities provided in
Ref.~\citenum{wichmann2007prediction} to test the accuracy of
our results in the range of molecular weights investigated in this
work ($30-160$~Da).  We systematically coarse-grained all the
compounds contained therein by means of the {\sc Auto-Martini}
tool \cite{Bereau2015}, and extracted those whose CG representation
consists of one and two beads. This set contains the small alcohols
(\emph{e.g.}, propanol and butanol) and the fully-unsubstituted
hydrocarbons discussed in the enhanced molecular design section of the
main text. A correlation plot of the $\log_{10} P$ predicted by
coarse-grained simulations with their experimental
$\log_{10}\text{BB}$ counterparts is presented in
Fig.~\ref{fig:corr_plot}.

 Given the limited range in
$\log_{10}\text{BB}$ covered by this database, we further decided to
include the subset of the small molecules already discussed in
Sec.~\ref{sec:correlation-at}, which in
Ref.~\citenum{carpenter2014method} were analyzed by means of atomistic
molecular dynamics simulations.  Overall, the permeability
coefficients extracted from CG simulations exhibit a high
correlation---$R^2\approx0.84$, with a mean absolute error of one
$\log_{10}$ unit in the permeability coefficient---with experimental
$\log_{10}\text{BB}$ measurements over a wide range of orders of
magnitude. 

In the case of chemical compounds mapping to Martini
unimers and dimers, we highlight the presence of sequence
of data points at constant $\log_{10} P$, spanning an interval of one
$\log_{10}\text{BB}$ unit on the vertical axis. This is due to the
reduction in chemical space generated by the transferable
coarse-grained model: from Fig.~\ref{fig:corr_plot} it is apparent
that chemically different compounds mapping to the same coarse-grained
representation exhibit similar permeation properties (in terms of
$\log_{10}\text{BB}$).

The accuracy of CG models when increasing the
molecular weight is again remarkable, with an exception in the case of
Salbutamol (``Salb'' in Fig.~\ref{fig:corr_plot}). The presence of
this outlier is connected to Martini parametrization issues for some
high molecular-weight compounds containing ring structures, a problem
currently under investigation.

\section{{\sc Gromacs} input files}
In this section we report the {\sc Gromacs} input files for the set of small molecules (extracted from Ref.~\citenum{carpenter2014method}) investigated by means of coarse-grained simulations: Atenolol, Cimetidine, Ibuprofen, Mannitol, Salbutamol and Salicylate.
\label{sec:gromacs}

\begin{Verbatim}
; GENERATED WITH auto_martini.py
; Atenolol
; Tristan Bereau (2014)

[moleculetype]
; molname       nrexcl
  MOL           2

[atoms]
; id    type    resnr   residue  atom    cgnr    charge  smiles
    1     P2      1     MOL       P01     1         0   ; CC(N)=O
    2     SC5     1     MOL       S01     2         0   ; c1ccccc1
    3     SC5     1     MOL       S02     3         0   ; c1ccccc1
    4     SC5     1     MOL       S03     4         0   ; c1ccccc1
    5     P2      1     MOL       P02     5         0   ; CO
    6     P3      1     MOL       P03     6         0   ; NCCO
    7     C2      1     MOL       C01     7         0   ; CCC

[bonds]
; i j     funct     length    force.c.
  1 2       1         0.25     1250
  4 5       1         0.37     1250
  5 6       1         0.26     1250
  6 7       1         0.24     1250

[constraints]
;  i   j     funct   length
   2   3     1       0.24
   2   4     1       0.24
   3   4     1       0.24

[angles]
; i j k         funct   angle   force.c.
  1 2 3         2       128.1   25.0
  1 2 4         2       143.7   25.0
  2 4 5         2       114.0   25.0
  3 4 5         2       54.2    25.0
  4 5 6         2       83.6    25.0
  5 6 7         2       117.2   25.0

\end{Verbatim}

\newpage
\begin{Verbatim}
; GENERATED WITH auto_martini.py
; Cimetidine
; Tristan Bereau (2014)

[moleculetype]
; molname       nrexcl
  MOL           2

[atoms]
; id    type    resnr   residue  atom    cgnr    charge  smiles
    1     SNd     1     MOL       S01     1         0   ; Cc1cncn1
    2     SP1     1     MOL       S02     2         0   ; c1cncn1
    3     N0      1     MOL       N01     3         0   ; CS
    4     C5      1     MOL       C01     4         0   ; CC
    5     P2      1     MOL       P01     5         0   ; CN=CN
    6     P2      1     MOL       P02     6         0   ; NC#N

[bonds]
; i j     funct     length    force.c.
  2 3       1         0.25     1250
  3 4       1         0.27     1250
  4 5       1         0.31     1250
  5 6       1         0.25     1250

[constraints]
;  i   j     funct   length
   1   2     1       0.22

[angles]
; i j k         funct   angle   force.c.
  1 2 3         2       65.5    25.0
  2 3 4         2       86.2    25.0
  3 4 5         2       152.0   25.0
  4 5 6         2       94.1    45.0

\end{Verbatim}

\newpage
\begin{Verbatim}
; GENERATED WITH auto_martini.py
; Ibuprofen
; Tristan Bereau (2014)

[moleculetype]
; molname       nrexcl
  MOL           2

[atoms]
; id    type    resnr   residue  atom    cgnr    charge  smiles
    1     C3      1     MOL       C01     1         0   ; CC(C)C
    2     SC5     1     MOL       S01     2         0   ; c1ccccc1
    3     SC5     1     MOL       S02     3         0   ; c1ccccc1
    4     SC5     1     MOL       S03     4         0   ; c1ccccc1
    5     Nda     1     MOL       N01     5         0   ; CCC(O)=O

[bonds]
; i j     funct     length    force.c.
  1 4       1         0.34     1250
  3 5       1         0.25     1250

[constraints]
;  i   j     funct   length
   2   3     1       0.24
   2   4     1       0.24
   3   4     1       0.24

[angles]
; i j k         funct   angle   force.c.
  1 4 2         2       69.2    25.0
  1 4 3         2       123.3   25.0
  2 3 5         2       144.4   25.0
  4 3 5         2       128.0   25.0

\end{Verbatim}

\newpage
\begin{Verbatim}
;;;; GENERATED WITH auto-martini
; Mannitol
; Tristan Bereau (2014)

[moleculetype]
; molname       nrexcl
  MOL           2

[atoms]
; id    type    resnr   residu  atom    cgnr    charge  smiles
  1     P4      1       MOL     P01     1       0     ; OCC(O)CO
  2     P4      1       MOL     P02     2       0     ; OCC(O)CO

[bonds]
; i j   funct   length  force.c.
  1 2   1       0.26    1250

\end{Verbatim}

\newpage
\begin{Verbatim}
; GENERATED WITH auto_martini.py
; Salbutamol
; Tristan Bereau (2014)

[moleculetype]
; molname       nrexcl
  MOL           2

[atoms]
; id    type    resnr   residue  atom    cgnr    charge  smiles
    1     P2      1     MOL       P01     1         0   ; CO
    2     SC5     1     MOL       S01     2         0   ; c1ccccc1
    3     SC5     1     MOL       S02     3         0   ; c1ccccc1
    4     SNda    1     MOL       S03     4         0   ; Oc1ccccc1
    5     P1      1     MOL       P02     5         0   ; CCO
    6     Nd      1     MOL       N01     6         0   ; CC(C)(C)N

[bonds]
; i j     funct     length    force.c.
  1 2       1         0.25     1250
  3 5       1         0.25     1250
  5 6       1         0.35     1250

[constraints]
;  i   j     funct   length
   2   3     1       0.24
   2   4     1       0.24
   3   4     1       0.24

[angles]
; i j k         funct   angle   force.c.
  1 2 3         2       121.3   25.0
  1 2 4         2       61.3    25.0
  2 3 5         2       61.3    25.0
  3 5 6         2       118.2   25.0
  4 3 5         2       121.5   25.0

\end{Verbatim}

\newpage
\begin{Verbatim}
; GENERATED WITH auto_martini.py
; Salicylate
; Tristan Bereau (2014)

[moleculetype]
; molname       nrexcl
  MOL           2

[atoms]
; id    type    resnr   residue  atom    cgnr    charge  smiles
    1     SC5     1     MOL       S01     1         0   ; c1ccccc1
    2     SNda    1     MOL       S02     2         0   ; Oc1ccccc1
    3     SC5     1     MOL       S03     3         0   ; c1ccccc1
    4     P1      1     MOL       P01     4         0   ; OC=O

[bonds]
; i j     funct     length    force.c.
  3 4       1         0.25     1250

[constraints]
;  i   j     funct   length
   1   2     1       0.24
   1   3     1       0.24
   2   3     1       0.24

[angles]
; i j k         funct   angle   force.c.
  1 3 4         2       122.0   25.0
  2 3 4         2       62.0    25.0

\end{Verbatim}

\newpage
\begin{Verbatim}
; GENERATED WITH auto_martini.py
; Benzoic acid
; Tristan Bereau (2014)

[moleculetype]
; molname       nrexcl
  MOL           2

[atoms]
; id    type    resnr   residue  atom    cgnr    charge  smiles
    1     P1      1     MOL       P01     1         0   ; OC=O
    2     SC5     1     MOL       S01     2         0   ; c1ccccc1
    3     SC5     1     MOL       S02     3         0   ; c1ccccc1
    4     SC5     1     MOL       S03     4         0   ; c1ccccc1

[bonds]
; i j     funct     length    force.c.
  1 4       1         0.25     1250

[constraints]
;  i   j     funct   length
   2   3     1       0.24
   2   4     1       0.24
   3   4     1       0.24

[angles]
; i j k         funct   angle   force.c.
  1 4 2         2       61.5    25.0
  1 4 3         2       121.5   25.0

\end{Verbatim}

\newpage
\begin{Verbatim}
; GENERATED WITH auto_martini.py
; Urea
; Tristan Bereau (2014)

[moleculetype]
; molname       nrexcl
  MOL           2

[atoms]
; id    type    resnr   residue  atom    cgnr    charge  smiles
    1     P4      1     MOL       P01     1         0   ; NC(N)=O

\end{Verbatim}

\clearpage

\section{Functional Group Distributions}

In this section we provide the statistics for the top five most
populous functional groups found in each highlighted region in Fig.~3
of the main text. For clarity, we reinsert a modified version of the figure 
below, with each region labeled with a number. The numbers correspond 
to entries in the tables detailing the total number of molecules and 
functional group populations found in each region. The instances 
of each functional group are detected using the {\sc checkmol} 
package~\cite{Haider2010}, which also provides the definition of each functional group in its
documentation. Note that we no longer color the points by their permeabilities, 
rather we use the number of heteroatom substitutions in each molecule. 
Zero corresponds to molecules made only of
carbons (saturated with hydrogens), while larger heteroatom
substitutions incorporate oxygens, nitrogens, and fluorines.
Unsurprisingly, we find that the number of heteroatom substitutions
acts as a good proxy for water/octanol partitioning, making the
compounds increasingly polar. 

\label{sec:fgroup-details}

\begin{figure*}[htbp]
  \begin{center}
    \includegraphics[width=0.9\linewidth]{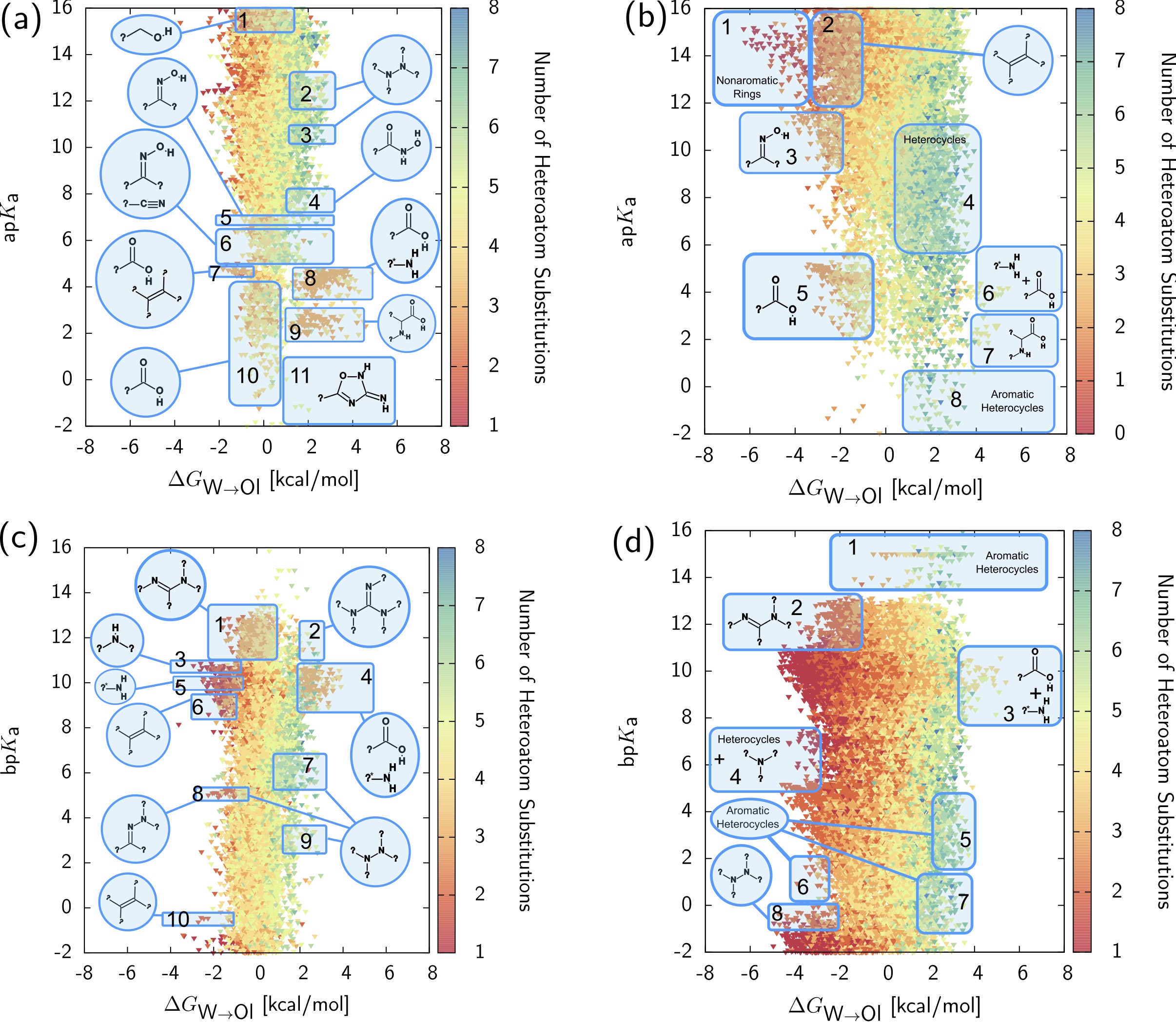}
    \caption{Adapation of Fig.~3 (main text) with each blue
      highlighted region labeled with a number.  Acidic and basic pKa
      are shown in panels (a,b) and (c,d), respectively. Panels (a,c)
      and (b,d) describe the coverage corresponding to coarse-grained
      unimers and dimers, respectively. The numbers of each region
      correspond to rows in the tables below that specify the
      percentage of the molecules containing specific functional
      groups. The points are now colored based on the number of
      heteroatom substitutions per molecule.}
    \label{fig:fgroupnumbered}
  \end{center}
\end{figure*}

\begin{table}[htbp]
  \caption{Table detailing top two functional group populations for regions highlighted in Fig.~\ref{fig:fgroupnumbered}a.}
  \label{tab:S1a}

  \begin{center}
    \begin{tabular}{ | c | c | c | c | }
      \hline
      \textbf{ Region \#} & \textbf{Total Molecules} &\textbf{ FG 1, \%} & \textbf{FG 2, \%}\\
      \hline 
      1  & 7231  & prim. alcohol, 58.5  & alkene, 38.3 \\
      \hline
      2  & 493 & hydrazine, 50.1  & nitrile, 21.3 \\
      \hline
      3  & 275 & hydrazine, 50.2  & carboxylic acid sec. amide, 21.5 \\
      \hline
      4  & 145  & hydroxamic acid, 51.0  & nitrile, 25.5 \\
      \hline
      5  & 367  & oxime, 57.0  & nitrile, 36.0 \\
      \hline
      6  & 788  & oxime, 53.7  & nitrile, 52.9  \\
      \hline
      7  & 90  & carboxylic acid, 78.9  & alkene, 53.3  \\
      \hline
      8  & 461  & carboxylic acid, 97.8  & prim. aliphat. amine, 72.7  \\
      \hline
      9  & 271  & carboxylic acid, 92.3  & alpha-aminoacid, 59.8  \\
      \hline
      10  & 1187  & carboxylic acid, 80.9  & alkyne, 36.5  \\
      \hline
      11  & 12  & iminohetarene, 91.7  & aromatic compound, 91.7  \\
      \hline
    \end{tabular}
  \end{center}
\end{table}

\begin{table}[htbp]
  \caption{Continuation of table S1, detailing top third through fifth functional group populations for regions highlighted in Fig.~\ref{fig:fgroupnumbered}a.}
  \label{tab:S1b}

  \begin{center}
    \begin{tabular}{ | c | c | c | c | }
      \hline
      \textbf{ Region \#} &\textbf{ FG 3, \%} & \textbf{FG 4, \%} & \textbf{FG 5, \%}\\
      \hline 
      1   & alkyne, 21.2  & hydrazine, 18.8  & nitrile, 16.1\\
      \hline
      2   & sec. alcohol, 20.9  & prim. alcohol, 17.6  & alkyne, 14.4\\
      \hline
      3  & nitrile, 20.0  & alkyne, 16.4  &   urea, 10.2\\
      \hline
      4    & hydrazine, 22.1  & aldehyde, 15.2  & oxime, 14.5\\
      \hline
      5    & alkyne, 30.0  & alkyl fluoride, 15.8  & ketone, 12.8\\
      \hline
      6    & alkyne, 27.3  & alkyl fluoride, 18.9  & aldehyde, 17.1\\
      \hline
      7    & alkyl fluoride, 25.6  & prim. alcohol, 14.4  & hydrazine, 13.3\\
      \hline
      8    & alkene, 58.8  & sec. aliphat. amine 24.5  & alkyne, 13.9\\
      \hline
      9   & prim. aliphat. amine, 49.1  & alkene, 39.1  & sec. aliphat. amine, 35.8\\
      \hline
      10   & nitrile, 28.1  & alkene, 25.9  & oxime, 14.3\\
      \hline
      11   & heterocyclic compound, 91.7  & aldehyde, 33.3  & alkyne, 33.3\\
      \hline
    \end{tabular}
  \end{center}
\end{table}

\begin{table}[htbp]
  \caption{Table detailing top two functional group populations for regions highlighted in Fig.~\ref{fig:fgroupnumbered}b.}
  \label{tab:S2a}

  \begin{center}
    \begin{tabular}{ | c | c | c | c | }
      \hline
      \textbf{ Region \#} & \textbf{Total Molecules} &\textbf{ FG 1, \%} & \textbf{FG 2, \%}\\
      \hline 
      1  & 201  & nonaromatic ring, 99.5  & alkene, 99.5 \\
      \hline
      2  & 3769 & alkene, 70.3  & alkyne, 42.1 \\
      \hline
      3  & 724 & oxime, 65.9  & alkene, 65.3 \\
      \hline
      4  & 17068  & heterocyclic compound, 49.6  & aromatic compound, 41.4 \\
      \hline
      5  & 1914  & carboxylic acid, 79.8  & alkene, 56.6 \\
      \hline
      6  & 215  & carboxylic acid, 83.3  & prim. aliphat. amine, 66.0  \\
      \hline
      7  & 172  & carboxylic acid, 85.5  & alpha-aminoacid, 69.2  \\
      \hline
      8  & 668  & aromatic compound, 97.9  & heterocyclice compound, 97.9  \\
      \hline
    \end{tabular}
  \end{center}
\end{table}

\begin{table}[htbp]
  \caption{Continuation of table S3, detailing top third through fifth functional group populations for regions highlighted in Fig.~\ref{fig:fgroupnumbered}b.}
  \label{tab:S2b}

  \begin{center}
    \begin{tabular}{ | c | c | c | c | }
      \hline
      \textbf{ Region \#} &\textbf{ FG 3, \%} & \textbf{FG 4, \%} & \textbf{FG 5, \%}\\
      \hline 
      1   & alkyne, 8.5  & halogen deriv., 8.0  & alkyl fluoride, 2.0\\
      \hline
      2   & alkyl fluoride, 25.3  & prim. alcohol, 16.5  & sec. alcohol, 16.1\\
      \hline
      3  & alkyne, 36.3  & alkyl fluoride, 14.8  &  heterocyclic compound, 9.9\\
      \hline
      4    & iminohetarene, 23.4  & oxohetarene, 18.5  & hydrazine, 18.4\\
      \hline
      5    & alkyl fluoride, 24.1  & alkyne, 20.2  & heterocyclic compound, 13.5\\
      \hline
      6    & sec. aliphat. amine, 30.2  & heterocyclic compound, 26.5  & aromatic compound, 17.2\\
      \hline
      7    & sec. aliphat. amine, 56.4  & prim. aliphat. amine, 51.7  & heterocyclic compound, 27.3\\
      \hline
      8    & iminohetarene, 96.3  & hydrazine 25.1  & phenol, 23.4\\
      \hline
    \end{tabular}
  \end{center}
\end{table}

\begin{table}[htbp]
  \caption{Table detailing top two functional group populations for regions highlighted in Fig.~\ref{fig:fgroupnumbered}c.}
  \label{tab:S3a}

  \begin{center}
    \begin{tabular}{ | c | c | c | c | }
      \hline
      \textbf{ Region \#} & \textbf{Total Molecules} &\textbf{ FG 1, \%} & \textbf{FG 2, \%}\\
      \hline 
      1  & 3651  & carboxylic acid amidine, 61.3  & alkene, 36.5 \\
      \hline
      2  & 66 & guanidine, 63.6  & hydrazine, 42.4 \\
      \hline
      3  & 356 & sec. aliphat. amine, 55.3  & alkene, 48.9 \\
      \hline
      4  & 778  & carboxylic acid, 68.5  & prim. aliphat. amine, 59.5 \\
      \hline
      5  & 636  & alkene, 66.8  & prim. aliphat. amine, 57.9 \\
      \hline
      6  & 626  & alkene, 63.4  & prim. aliphat. amine, 48.9  \\
      \hline
      7  & 2439  &  hydrazine, 61.3  & nitrile, 19.0  \\
      \hline
      8  & 123  & hydrazine, 75.6  & hydrazone, 60.2  \\
      \hline
      9  & 302  & hydrazine, 74.8  & carboxylic acid hydrazide, 26.5  \\
      \hline
      10  & 32  & hydrazine, 62.5  & alkene, 56.3  \\
      \hline
    \end{tabular}
  \end{center}
\end{table}

\begin{table}[htbp]
  \caption{Continuation of table S5, detailing top third through fifth functional group populations for regions highlighted in Fig.~\ref{fig:fgroupnumbered}c.}
  \label{tab:S3b}

  \begin{center}
    \begin{tabular}{ | c | c | c | c | }
      \hline
      \textbf{ Region \#} &\textbf{ FG 3, \%} & \textbf{FG 4, \%} & \textbf{FG 5, \%}\\
      \hline 
      1   & enamine, 22.5  & guanidine, 20.1  & hydrazine, 16.6\\
      \hline
      2   & carboxylic acid amidine, 19.7  & carboxylic acid, 16.7  & prim. alcohol, 13.6\\
      \hline
      3  & prim. aliphat. amine, 44.1  & alkyl fluoride, 24.7  &   alkyne, 11.8\\
      \hline
      4    & alkene, 40.62  & sec. aliphat. amine, 17.7  & hydrazine, 16.2\\
      \hline
      5    & sec. aliphat. amine, 34.1  & alkyl fluoride, 29.7  & alkyne, 16.8\\
      \hline
      6    & alkyne, 33.4  & sec. aliphat. amine, 30.2  & alkyl fluoride, 27.8\\
      \hline
      7    & alkyne, 15.9  & carboxylic acid amidrazone, 10.0  & prim. alcohol, 10.0\\
      \hline
      8    & alkene, 43.1  & alkyne 18.7  & nitrile, 9.8\\
      \hline
      9   & semicarbazide, 22.8  & prim. alcohol, 14.6  & alkyne, 13.9\\
      \hline
      10   & alkyne, 25.0  & nitrile, 12.5  & oxime, 9.4\\
      \hline
    \end{tabular}
  \end{center}
\end{table}

\begin{table}[htbp]
  \caption{Table detailing top two functional group populations for regions highlighted in Fig.~\ref{fig:fgroupnumbered}d.}
  \label{tab:S4a}

  \begin{center}
    \begin{tabular}{ | c | c | c | c | }
      \hline
      \textbf{ Region \#} & \textbf{Total Molecules} &\textbf{ FG 1, \%} & \textbf{FG 2, \%}\\
      \hline 
      1  & 768  & heterocyclic compound, 90.5  & aromatic compound, 88.3 \\
      \hline
      2  & 3599 & carboxylic acid amidine, 71.3  & alkene, 53.0 \\
      \hline
      3  & 688 & prim. aliphat. amine, 82.3  & carboxylic acid, 72.1 \\
      \hline
      4  & 778  & heterocyclic compound, 94.9  & tert. aliphat. amine, 75.9 \\
      \hline
      5  & 1508  & heterocyclic compound, 69.5  & aromatic compound, 63.3 \\
      \hline
      6  & 91  & aromatic compound, 76.9  & heterocyclic compound, 76.9  \\
      \hline
      7  & 2528  &  heterocyclic compound, 65.4  & aromatic compound, 57.0  \\
      \hline
      8  & 291  & hydrazine, 68.0  & alkene, 58.1  \\
      \hline
    \end{tabular}
  \end{center}
\end{table}

\begin{table}[htbp]
  \caption{Continuation of table S7, detailing top third through fifth functional group populations for regions highlighted in Fig.~\ref{fig:fgroupnumbered}d.}
  \label{tab:S4b}

  \begin{center}
    \begin{tabular}{ | c | c | c | c | }
      \hline
      \textbf{ Region \#} &\textbf{ FG 3, \%} & \textbf{FG 4, \%} & \textbf{FG 5, \%}\\
      \hline 
      1   & iminohetarene, 86.1  & phenol, 27.6  & hydrazine, 15.6\\
      \hline
      2   & enamine, 18.8  & heterocyclic compound, 17.0  & alkyl fluoride, 11.0\\
      \hline
      3  & alpha-aminoacid, 28.5  & sec. aliphat. amine, 23.8  &   alkene, 15.6\\
      \hline
      4    & alkene, 55.7  & alkyne, 21.7  & sec. aliphat. amine, 16.1\\
      \hline
      5    & hydrazine, 50.1  & oxohetarene, 37.7  & iminohetarene, 34.8\\
      \hline
      6    & alkene, 52.7  & alkyne, 22.0  & oxime ether, 8.8\\
      \hline
      7    & iminohetarene, 34.8  & oxohetarene, 31.8  & hydrazine, 29.6\\
      \hline
      8    & tert. alcohol, 22.7  & alkyne 10.0  & sec. alcohol, 6.5\\
      \hline
    \end{tabular}
  \end{center}
\end{table}

\clearpage

\bibliographystyle{bibliography}
\bibliography{biblio_supp}